\documentclass[usenatbib,twocolumn]{mn2e}
\usepackage{times}
\usepackage{graphicx}
\usepackage{amsfonts}
\usepackage{epsfig,rotating,natbib,pstricks}
\usepackage{color}
\usepackage{amssymb,latexsym}
\usepackage{amsmath}

\pubyear{2010}

\def\beq{\begin{equation}}
\def\eeq{\end{equation}}   
\def\baq{\begin{eqnarray}}
\def\eaq{\end{eqnarray}}

\def\p3m{P$^3$M}
\def\ap3m{AP$^3$M}

\def\h1{H\/I}
\def\omegah1{\Omega_{\h1}}
\def\ph1{P_{_{\h1}}}
\def\ph1k{P_{_{\h1}}(k)}
\def\dh1k{\Delta^2_{_{\h1}}(k)}

\def\msun{\mathrm{M}_{\odot}}

\newcommand{\be}{\begin{equation}}
\newcommand{\e}{\end{equation}}

\title[\h1 at z $\simeq 1$ ]{Detecting neutral hydrogen in emission at redshift z $\simeq$ 1}

\author[Khandai et al.]{\parbox{18cm}{
Nishikanta Khandai$^{1}$, 
Shiv K. Sethi$^{1,2}$, 
Tiziana Di Matteo$^{1}$, 
Rupert A.C. Croft$^{1}$, 
Volker Springel$^{3,4}$, 
Anirban Jana$^5$, 
Jeffrey P. Gardner$^6$}\vspace{0.3cm}\\
$^1$ {McWilliams Center for Cosmology, 
  Carnegie Mellon University, 5000 Forbes Avenue, Pittsburgh, PA 15213, USA}\\
$^2$ {Raman Research Institute, C. V. Raman Avenue, Sadashivanagar, Bangalore 560080, India}\\
$^3$ {Heidelberg Institute for Theoretical Studies, Schloss-Wolfsbrunnenweg 35, 69118 Heidelberg, Germany} \\
$^4$ {Zentrum f\"{u}r Astronomie der Universit\"{a}t Heidelberg,
    Astronomisches Recheninstitut, M\"{o}nchhofstr. 12-14, 69120 Heidelberg, Germany}\\
$^5$ {Pittsburgh Supercomputing Center, 300 S. Craig Street, Pittsburgh, PA 15213, USA} \\
$^6$ {University of Washington, Department of Physics, Seattle, WA 98195-1560, USA}
}
\pubyear{2010}

\label{firstpage}

\def\LaTeX{L\kern-.36em\raise.3ex\hbox{a}\kern-.15em
    T\kern-.1667em\lower.7ex\hbox{E}\kern-.125emX}

\pagerange{\pageref{firstpage}--\pageref{lastpage}} 

\begin{document}

\maketitle

\begin{abstract}
  We use a large N-body simulation to examine the detectability of HI in
  emission at redshift $z \simeq 1$, and the constraints imposed by
  current observations on the neutral hydrogen mass function of galaxies
  at this epoch. We consider three different models for populating dark
  matter halos with HI, designed to encompass uncertainties at this
  redshift.  These models are consistent with recent observations of the
  detection of HI in emission at $z\simeq 0.8$.  Whilst detection of
  $21\,{\rm cm}$ emission from individual halos requires extremely long
  integrations with existing radio interferometers, such as the Giant
  Meter Radio Telescope (GMRT), we show that the stacked $21\,{\rm cm}$
  signal from a large number of halos can be easily detected. However,
  the stacking procedure requires accurate redshifts of galaxies. We
  show that radio observations of the field of the DEEP2 spectroscopic
  galaxy redshift survey should allow detection of the HI mass function
  at the $5\hbox{--}12\sigma$ level in the mass range
  $10^{11.4}h^{-1}{\rm \msun}\leq M_{\mathrm{halo}}\leq
  10^{12.5}h^{-1}{\rm \msun}$, with a moderate amount of observation
  time.  Assuming a larger noise level that corresponds to an upper
  bound for the expected noise for the GMRT, the detection significance
  for the HI mass function is still at the $1.7\hbox{--}3 \sigma$ level.
  We find that optically undetected satellite galaxies enhance the HI
  emission profile of the parent halo, leading to broader wings as well
  as a higher peak signal in the stacked profile of a large number of
  halos.  We show that it is in principle possible to discern the
  contribution of undetected satellites to the total HI signal, even
  though cosmic variance limitation make this challenging for some of
  our models.
\end{abstract}



\section{Introduction}

Observations show that the cosmic star formation rate (SFR) has declined
by more than an order of magnitude since $z \simeq 1$
\citep{2004ApJ...615..209H}.  However, a combined census of the cold
gas, the fuel for star formation, and stellar components is still
largely missing in observations.  The cold gas fraction of a halo is a
crucial ingredient in models of galaxy formation and constitutes the
link to how galaxies obtain gas and subsequently convert it to
stars. Hence, measurements of HI in the post-reionization era can place
tight constraints on different models of galaxy formation
\citep{2009astro2010S.241P}.

After the epoch of reionization, the neutral hydrogen (HI) survives in
dense clouds, e.g.~damped Lyman-$\alpha$ systems (DLAs) and Lyman-limit
systems (LLS), that are high-redshift equivalents of the HI-rich
galaxies that we see at the present epoch.  The baryon fraction locked
up in HI, $\Omega_{\rm HI}$, in star-forming galaxies in the
post-reionization epoch can be determined from the study of damped
Lyman-$\alpha$ systems in absorption for $0.5 \le z \le 5$
\citep{2005ApJ...635..123P,2006ApJ...636..610R,2009A&A...505.1087N}.
Even though these observations give clues about aggregate behaviour of
star formation as a function of redshift, they cannot be used to infer
the total HI mass of these systems because they are seen in absorption.
At $z\simeq 1$, even the detection of HI in damped Lyman-$\alpha$ has
not been easy as the Lyman-$\alpha$ frequency is not accessible to
ground-based telescopes.  At this redshift, constraints on the global HI
fraction come from associated $\rm MgII$ systems, HST observations
\citep[for details see][and references therein]{2006ApJ...636..610R},
and the absorption of 21~cm radiation from bright background radio
sources \citep{2009MNRAS.396..385K}, but with significant uncertainties
on the estimated  HI fraction.  Direct observation of HI in
emission and its detailed modelling has only been possible at $z \simeq
0$ thus far \citep{2005MNRAS.359L..30Z}.

Direct observation in 21cm emission of ten massive galaxies have 
been reported for $0.17 < z < 0.25 $, with the Arecibo telescope \citep{2008ApJ...685L..13C}.
At higher redshifts, the HI emission from individual clouds is too weak to be
detectable with present radio instruments 
\citep{2010MNRAS.407..567B}. 
A long integration time is required for
detecting even the brightest objects since the peak signal is a few tens
of micro Jansky whereas the system noise is of the order of hundreds of
micro Jansky.  
A possible approach to circumvent the difficulty of
detecting individual clouds lies in stacking the HI emission of galaxies
with known redshifts.  This approach has been attempted for both cluster
galaxies and the field galaxies in the recent past
\citep{2007MNRAS.376.1357L,2009MNRAS.399.1447L}.  In particular, a
similar line of study has resulted in the recent detection of HI at $z
\simeq 0.8$ \citep{2010Natur.466..463C}.  An alternative approach rests
on the possible detection of the fluctuation in the redshifted HI
emission from high redshifts
\citep{2001JApA...22..293B,2008PhRvL.100i1303C,2009PhRvD..79h3538B}.

On the theoretical side, semi-analytical models of galaxy formation 
have looked at the evolution of cold gas (both in atomic and molecular form)
in galaxies and their results match with observations at $z =0$ 
\citep{2009ApJ...696L.129O,2009ApJ...698.1467O,2009ApJ...703.1890O,2009MNRAS.400..665O,
2010MNRAS.406...43P,2010MNRAS.409..515F,2010arXiv1003.0008K}.
However observations at higher redshifts are needed to better constrain 
the evolution of cold gas predicted by these models.

Given the importance of connecting cold gas and stars at $z \simeq 1$
over a wide range of galaxy environments, it is crucial to make
predictions for various detection strategies for current and upcoming
telescopes.  In this work we focus on the stacking method of individual
galaxies with known redshifts to predict how well the HI mass function
at $z = 1$ can be constrained with existing surveys and telescopes (in
particular the DEEP2\footnote{http://deep.berkeley.edu} survey and the
GMRT\footnote{http://gmrt.ncra.tifr.res.in}; but note that our method
is generic and can be extended to future surveys and instruments).  By
stacking we can also study the contribution of small satellite galaxies,
which are undetected in an optical survey but (as we shall show) contain
non-negligible amounts of HI, to the total $21\,{\rm cm}$ signal in
emission and also examine the constraints that one can put on the HI
mass function.

We model the HI in dark matter halos in a large $N-$body simulation by
refining the model of \cite{2010MNRAS.407..567B}.  Given the paucity of
observations at the redshifts under consideration, and our limited
understanding of how HI populates dark matter halos at these redshifts,
we consider a variety of models.  These are constrained by observations
of HI at low redshift, simulations of DLAs in small-volumes at high
redshift, as well as by some results of semi-analytical models of galaxy
formation at intermediate redshifts.  In particular, the models that we
consider are consistent with recent observations of HI in emission at $z
\simeq 0.8$ \citep{2010Natur.466..463C}.  

Our paper is organised as follows. We present our large dark matter
simulation in Section~\ref{sec_nbody}, and describe our model for the HI
distribution in the simulation along-with specifications of the DEEP2
survey as well as the GMRT in Section~\ref{sec_h1model}.  We discuss our
stacking procedure of individual galaxies and the contribution of
undetected satellites to the stacked HI spectra in
Section~\ref{sec_h1signal}.  In Section~\ref{sec_results}, we present
our results and discuss the prospects of detection with the GMRT, and
the constraints that one can put on the HI mass function.  We revisit
the issue of undetected satellites and its effect on the HI mass
function and discuss whether their presence can be detected. Finally, we
present our conclusions in Section~\ref{sec_conclusions}.


\begin{table}
      \begin{center}
        \begin{tabular}{c|c|c|c}
          \hline
          $L_{\mathrm{box}}$ & $N_{\mathrm{part}}$ & $m_{_{\mathrm{DM}}}$ & $\epsilon$ \\
          $\left(h^{-1}\mathrm{Mpc}\right)$& & $\left(10^8 h^{-1}\mathrm{M}_{\odot}\right)$ &  $\left(h^{-1}\mathrm{kpc}\right)$ \\
          \hline
          400 & $2448^3$ & 3.1 & 6.5  \\
         \hline
        \end{tabular}
      \end{center}
\caption{Basic simulation parameters for our dark matter run. The columns
  list the size of the simulation box, $L_{\mathrm{box}}$, the
  number of dark matter particles used in the simulation,
  $N_{\mathrm{part}}$, the mass of a single dark matter particle,
  $m_{_{\mathrm{DM}}}$, and the gravitational softening length,
  $\epsilon$.  All length scales are in comoving units.}
\label{table_simparam}
\end{table}

\section{N-body Simulation}
\label{sec_nbody}
We have used {\small P-GADGET}, a significantly upgraded version of
{\small GADGET2} \citep{2005MNRAS.364.1105S} which we are developing for
upcoming Petascale supercomputer facilities, for running a large dark
matter (DM) simulation in a $\Lambda$CDM cosmology.  The cosmological
parameters used were $\sigma_8 = 0.8$, $n_s=0.96$,
$\Omega_{\Lambda}=0.74$, and $\Omega_{\mathrm{m}}=0.26$.  The initial
conditions were generated with the Eisenstein and Hu power spectrum at
an initial redshift of $z=159$.  Table~\ref{table_simparam} lists the
basic simulation parameters: the size of the box $L_{\mathrm{box}}$, the
number of particles $N_{\mathrm{part}}$, the mass of a dark matter
particle $m_{_{\mathrm{DM}}}$ and the softening length $\epsilon$.  Note
for reference, our simulation volume is roughly half that of the
Millennium Simulation \citep{2005Natur.435..629S} but our mass
resolution is about a factor of three better.

The frequency and redshift widths corresponding to $L_{\mathrm{box}} =
400 \mathrm{h}^{-1} \mathrm{Mpc}$ at $z = 1$ are
$\Delta\nu_{\mathrm{box}} = 75.8$MHz and $\Delta z_{\mathrm{box}} =
0.239$.  The high resolution and large volume of our simulation enables
us to resolve the smallest groups expected to host HI, as well as to
look for effects of cosmic variance on observables like the HI mass
function.  Furthermore, we are able to resolve subhalos in the larger
halos.  In fact, we will use the distribution of subhalos in redshift
space to make predictions on how these subhalos affect the total HI
signal.  

We use the {\small SUBFIND} code \citep{2001MNRAS.328..726S} to find the
subhalo catalogue and to measure properties like central coordinate,
peculiar velocity, bound mass, maximum circular velocity and velocity
dispersion for every subhalo.  Groups of particles are retained as a
subhalo when they have at least 20 bound particles, which corresponds to
a minimum group mass of $M_{\mathrm{halo}} = 6.3\times 10^9
h^{-1}\mathrm{M}_{\odot}$.  This mass is slightly larger than the mass
of the smallest halo which is capable to host HI, as discussed in
Section~\ref{sec_h1model}.  The largest subhalo in an FOF halo is
generally characterised by {\small SUBFIND} as the central halo, and the
other bound structures as satellites. Since the central halo contains
most of the mass of the halo, we will loosely refer to it as the halo,
and to the smaller ones in its vicinity as subhalos or satellites, where
appropriate.

\section{Modelling The HI Distribution}
\label{sec_h1model}

\begin{figure*}
\begin{tabular}{cc}
  \includegraphics[width=3.3truein]{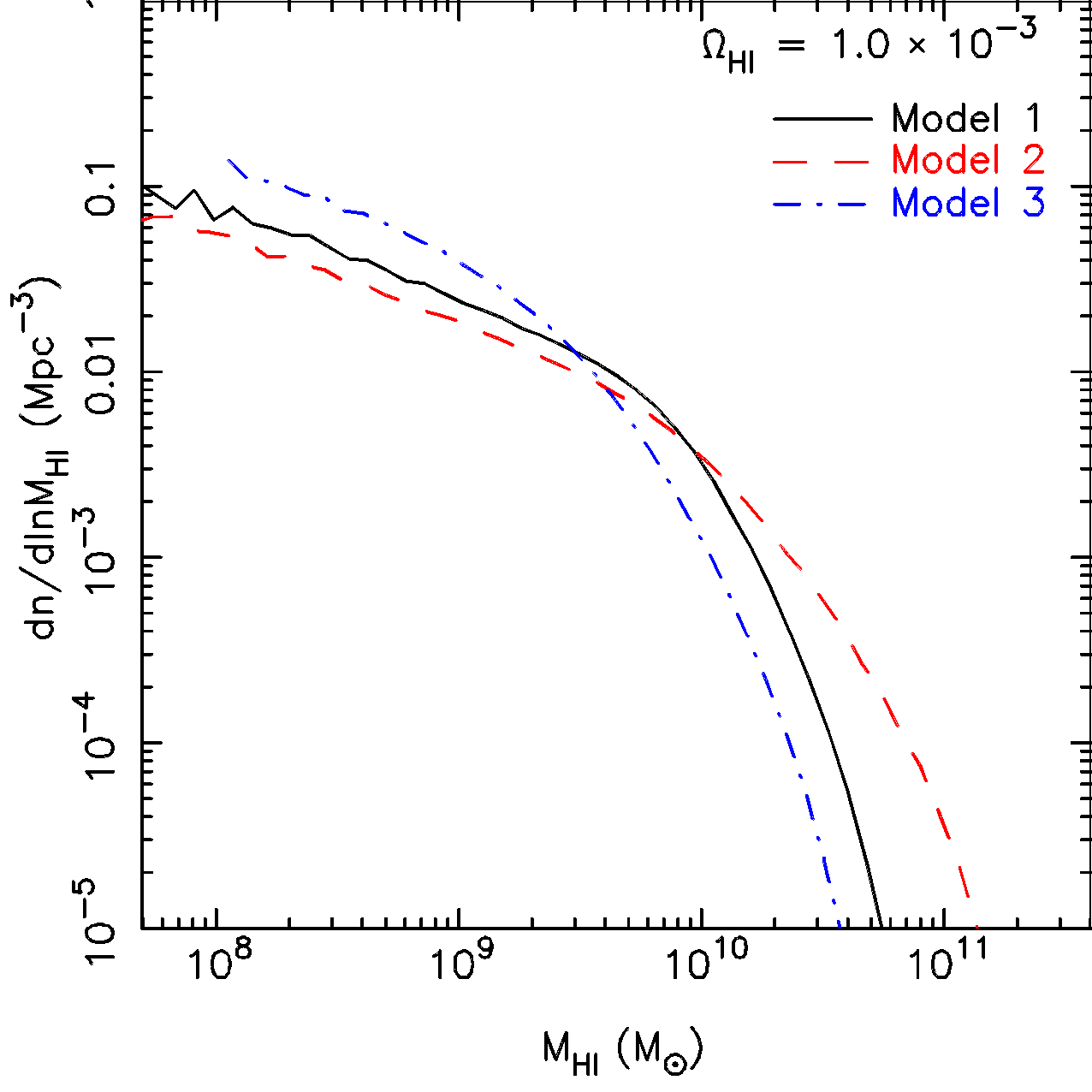} 
  \includegraphics[width=3.3truein]{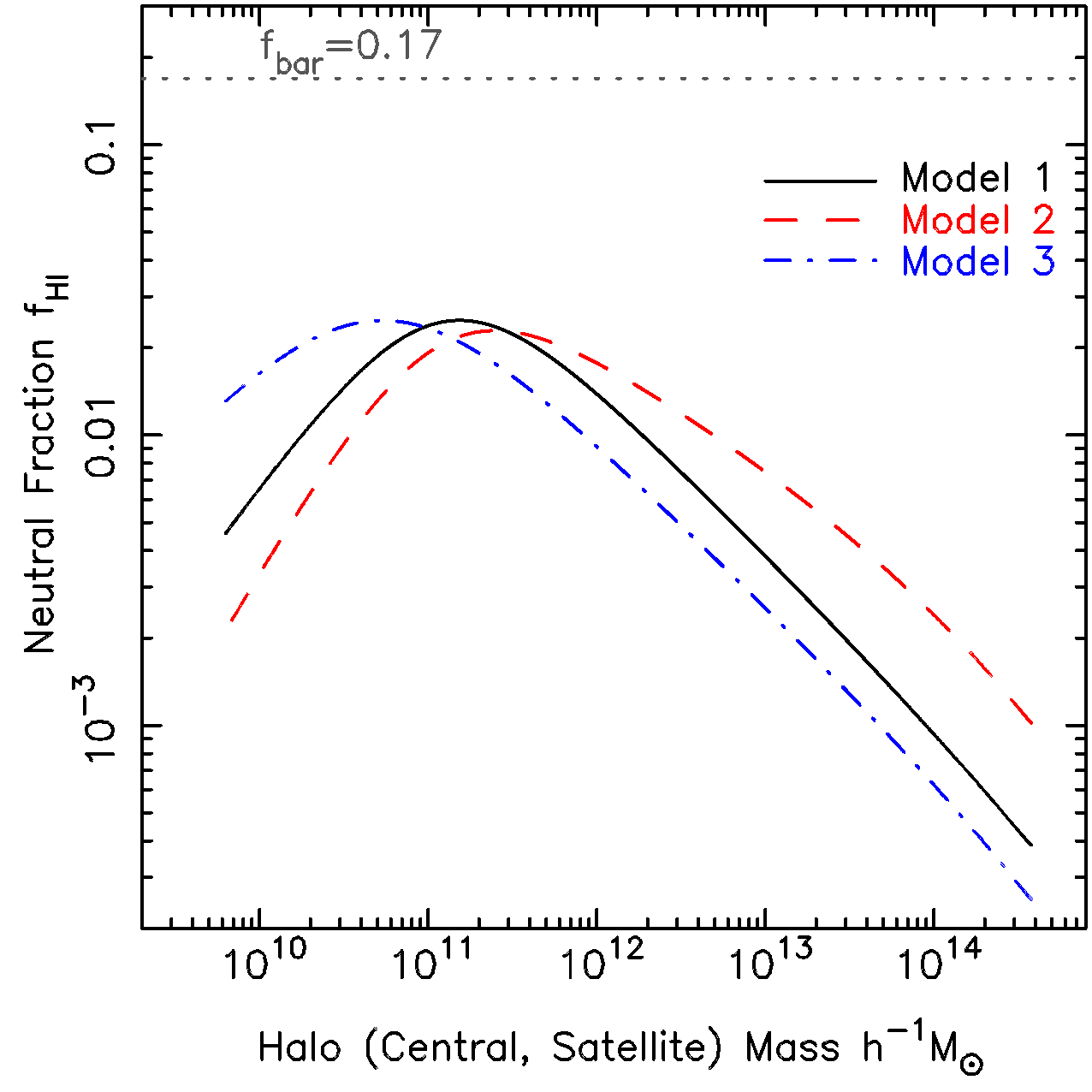} 
\end{tabular}
\caption{{\em Left:} The mass function for the three models that we
  consider.  Model~1 (solid line) is the \textcolor{black}{Zwaan et
    al. (2005)} mass function but normalised to $\Omega_{\mathrm{HI}} =
  10^{-3}$. Models~2 (dashed) and 3 (dot-dashed) are variations around
  Model~1, see Table~\ref{table_MHI2M} and Eqn.~(\ref{eq_MHI_M}) for
  details of the model parameters.  {\em Right:} The neutral mass
  fraction $f_{\mathrm{HI}} = M_{\mathrm{HI}}/ M_{\mathrm{halo}}$ as a
  function of halo mass for the three models.  The (dotted) horizontal
  line is the baryon mass fraction.}
\label{fig_massfn}
\end{figure*}
 
Our knowledge of the \h1 distribution in the Universe out to $z \simeq
5$ is derived mainly from QSO absorption spectra, where the gas absorbs
in the Lyman-$\alpha$ transition of the hydrogen atom.  We know from
observations that much of the inter-galactic medium (IGM) is highly
ionized and does not contain a significant amount of neutral hydrogen.
Instead, most of the neutral hydrogen resides in relatively rare damped
Lyman-$\alpha$ systems \citep{2005ARA&A..43..861W}.  DLAs and other high
column density absorption features are believed to arise due to gas
within galaxies \citep{2000ApJ...534..594H,2001ApJ...559..131G}.  It is
possible to make an estimate of the total neutral hydrogen content in
DLAs and study the evolution of the total neutral hydrogen content of
the Universe \citep{1996MNRAS.283L..79S, 2000ApJS..130....1R,
  2005MNRAS.363..479P,2005ApJ...635..123P,2006ApJ...636..610R,
  2009A&A...505.1087N}. Interestingly, these observations suggest that the
neutral hydrogen content of the Universe is almost constant in the
redshift range $0.5 \leq z \leq 5$, with a density parameter of
$\Omega_{\rm \h1} \simeq 0.001$.

At low redshifts, the \h1 content can be estimated more directly through
emission in the hyperfine transition.  Observations of the
HIPASS\footnote{HI Parkes All Sky Survey:
  http://www.parkes.atnf.csiro.au} galaxies \citep{2005MNRAS.359L..30Z}
in the local universe indicate a much lower neutral hydrogen content
$\left(\Omega_{\mathrm{HI}}(z=0) \simeq 4.6\times10^{-4}\right)$ than
seen at $ z \geq 1$.  These authors observed HI in emission of
$\simeq 4000$ galaxies in the local Universe to estimate the HI mass
function.  At higher redshifts, observations in the DEEP2 survey
\citep{2007MNRAS.376.1425G} indicate that the fraction $f_b$ of blue
galaxies (generally associated with late type gas-rich galaxies with
significant star formation activity) in groups is much higher at
redshifts $0.75 \leq z \leq 1.3$, increasing from $f_b = 0.84$ at $z =
0.75$ to $f_b = 0.94$ at $z = 1.3$, than what is observed in the local
Universe. These observations also suggest that $f_b$ for group and field
galaxies approaches the same value by $z=1.3$.  We use these observations
to motivate our model of assigning HI to dark matter halos at $z = 1$.

Here we use the observations of HI in emission at $z = 0$
\citep{2005MNRAS.359L..30Z} to match the HI mass function to the dark
matter halo mass function. We remind the reader that the halo catalogue
consists of both centrals and satellites.  The
\citet{2005MNRAS.359L..30Z} HI mass function is given in a
Schechter-like form: \beq \Theta\left(M_{\mathrm{HI}}\right)=
\theta^*\left(\frac{M_{\mathrm{HI}}}{M_{\mathrm{HI}}^*}\right)^{-\alpha}\exp
\left(-\frac{M_{\mathrm{HI}}}{M_{\mathrm{HI}}^*}\right)
\label{eq_zwaan}
\eeq where $\theta^*=6\times 10^{-3}h_{75}^3\mathrm{Mpc}^{-3}$ is the
normalisation factor, $\log(M_{\mathrm{HI}}^*/\msun)=9.8 h_{75}^{-2}$ is
the characteristic mass that defines the kink in the function, and
$\alpha=1.37$ is the slope at the low mass end.  $h_{75} = 0.75$ is the
dimensionless Hubble constant.  We vary the other models around it.

\cite{2010ApJ...718..972M} took a similar approach to compute the HI
bias out to redshifts $z = 4$.  They also incorporated the fraction of
blue galaxies in the local Universe, which is much smaller than what is
seen at $z = 1$ in their model.  For this study we take this fraction to
be unity.  Additionally, we use some input from semi-analytical models
and simulations of high-redshift DLAs to motivate our model.
Semi-analytical models \citep{2010MNRAS.406...43P,2010arXiv1003.0008K}
suggest that the shape of the HI mass function does not evolve
considerably, but shifts toward the high mass end with redshift, assuming 
a constant molecular to atomic hydrogen ratio, $\mathrm{H}_2/\mathrm{HI}$; though these results may change if 
this ratio is not a constant \citep{2009ApJ...696L.129O,2009ApJ...698.1467O,2009ApJ...703.1890O,2009MNRAS.400..665O}.  
This shift may be due to the higher HI content at high redshift, e.g.
$\Omega_{\mathrm{HI}}(z=1) \simeq 10^{-3} $ as compared to
$\Omega_{\mathrm{HI}}(z=0) \simeq 4.6\times10^{-4} $.  However, these
models do not match the low-end of the Zwaan et al. mass function, due
to finite resolution effects of their merger trees.  

Hydrodynamic simulations of DLAs  at $z=3$ by \citet{2008MNRAS.390.1349P}
yield a mapping from halo mass to HI mass, which can be described  by
\beq M_{\mathrm{HI}} \propto
\frac{\left(M_{_{\mathrm{halo}}}\right)^m}{1 +
  \left(\frac{M_{_{\mathrm{halo}}}}{M_{_{\mathrm{min}}}}\right)^n
  +
  \left(\frac{M_{_{\mathrm{halo}}}}{M_{_{\mathrm{max}}}}\right)^p
}.
\label{eq_MHI_M}
\eeq These authors found that there is a tight monotonic relation
between the virial mass and the HI mass of halos with some scatter.
They further found that the $M_{\mathrm{HI}} - M_{\mathrm{halo}}$
relation has a break at $M_{\mathrm{halo}}/\msun \simeq 10^{10.5} $,
suppressing HI in halos larger than this mass and a still stronger
suppression in halos with mass $M_{\mathrm{halo}}/\msun > 10^{11.0} $.
Furthermore, halos with masses as low as $M_{\mathrm{halo}}/\msun \simeq
10^{9.0}$ (or circular velocity at $z=3$ of $v_{\mathrm{circ}} \simeq
30\,{\rm km\,s^{-1}}$) are able to host a significant amount of HI.  The
gas in these halos is able to self-shield from the photo-ionising UV
background and maintain a significant amount of \h1 even though the
amount of gas is insufficient for sustaining star formation.

The form of Eq.~(\ref{eq_MHI_M}) contains two mass parameters,
$M_{\mathrm{min}}$ and $M_{\mathrm{max}}$, for the three regimes in the
$M_{\mathrm{HI}} - M_{\mathrm{halo}}$ relation of
\cite{2008MNRAS.390.1349P}.  Based on their simulations, we choose the
cutoff mass for halos not hosting any HI to be $M \simeq 10^{9.0}
h^{-1}\msun$ or $v_{\mathrm{circ}} \simeq 30\,{\rm km \,
  s^{-1}}$ at $z = 3$.  We use the scaling relation 
\begin{equation}
M_{\rm vir} \simeq 10^{10} ~{\mathrm M}_\odot \left( \frac{v_{\mathrm{circ}}}{60\, {\rm
      km \, s^{-1}}}\right)^3 \left(\frac{1 + z}{4}\right)^{-3/2} 
\label{eq_scaling}
\end{equation}
with $v_{\mathrm{circ}} = 30\,{\rm{km\,s^{-1}}}$ to determine the cutoff
mass of halos which do not host significant HI.  This translates to
${M}_{\mathrm{cutoff}}^{\rm halo}/\msun = 10^{9.55}$ at $z = 1$.

\begin{table}
      \begin{center}
        \begin{tabular}{c|c|c|c|c|c}
          \hline
          Model & $M_{\mathrm{min}}$ & $M_{\mathrm{max}}$& m & n & p\\
           & $\left(10^{10}h^{-1}{\rm M}_{\odot}\right)$ &  $\left(10^{10}h^{-1}\mathrm{M}_{\odot}\right)$&  &  & \\
          \hline
          1 & 12 & 143 & 1.8 & 1.36  & 1.8 \\
          2 & 12 & 143 & 2.0 & 1.36 & 2.0 \\
          3 & 5 & 143 & 1.6 & 1.15 & 1.7 \\
         \hline
        \end{tabular}
      \end{center}
\caption{Model Parameters: Mapping of HI mass to halo (centrals and
  satellites) mass for the three models that we consider.  See
  Eqn.~(\ref{eq_MHI_M}) for the functional form of the mapping from HI
  mass to halo mass.}
\label{table_MHI2M}
\end{table}

\begin{figure*}
\begin{tabular}{cc}
  \includegraphics[width=5.0truein]{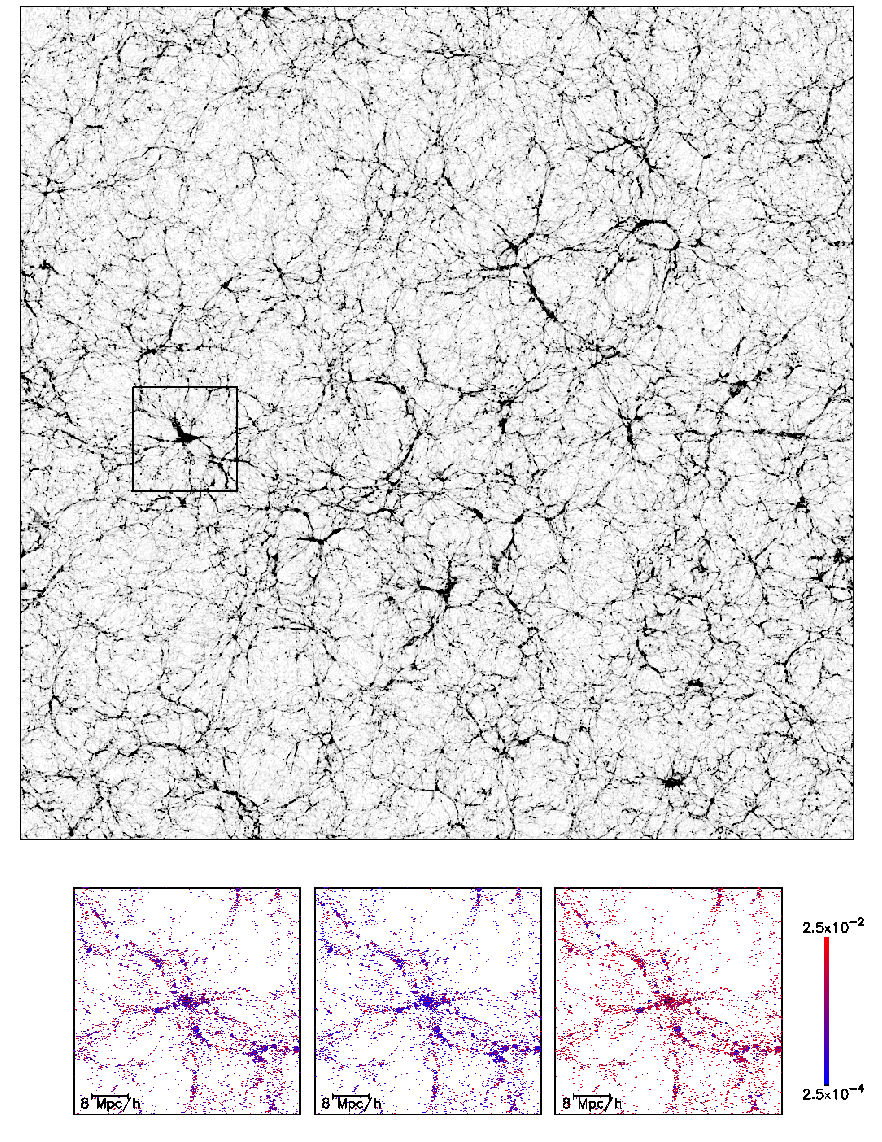} 
\end{tabular}
\caption{{\em Top:} A thin slice of our simulation enclosing the largest
  halo (square box), showing the distribution of dark matter particles.
  {\em Bottom:} A zoom-in for a region of dimension $50 \times 50 \times
  4 (h^{-1}{\rm Mpc})^3$ centred on the largest halo in our simulation,
  showing the halos which host HI, colour-coded with the neutral fraction
  $f_{\mathrm{HI}}$.  The panels from left to right are for the three
  models.}
\label{fig_slice}
\end{figure*}

Table~\ref{table_MHI2M} summarizes all the parameters for our three
models for the HI distribution over halos.  Our reference model
(model~1) matches the Zwaan et al. mass function but is renormalised to
$\Omega_{\mathrm{HI}}(z=1) = 10^{-3}$.  We also consider two alternative
models around the reference model.  In model 2, we allow a larger
fraction of HI in large mass halos and suppress HI in lower mass
halos. The third model is one in which the HI content in high mass halos
is suppressed and the HI is redistributed to lower mass halos.  Given
our lack of knowledge about how HI populates dark matter halos, these
three models should encompass a reasonable range of possibilities.  All
models are normalised to the fiducial value of $\Omega_{\mathrm{HI}} =
10^{-3}$.  The form of the mapping from dark matter halo mass,
$M_{\mathrm{halo}}$, to HI mass, $M_{\mathrm{HI}}$, is given in
Eq.~(\ref{eq_MHI_M}), which is a more generalised form of the mapping
considered by \cite{2010MNRAS.404..876W}.  Note that the ratio of the
indices \emph{m} and \emph{p} determines the HI content of halos with
mass $M_{\mathrm{halo}} > M_{\mathrm{max}}$.  In models 2 and 3, $m/p
= 1$, which means that the HI content in halos larger than
$M_{\mathrm{max}}$ approaches a constant.  On the other hand, the value
$m/p < 1$ for model 3 suppresses HI in larger halos.

The model mass functions for all the three models are shown in the
left panel of Figure~\ref{fig_massfn}.  The fiducial mass function (solid
line) is the one whose shape matches that of the Zwaan et. al mass function,
but is normalised to $\Omega_{\mathrm{HI}} = 10^{-3}$. The mass function of
model~2 (dashed line) has comparatively more HI in larger halos while the
HI in smaller halos is suppressed.  On the other hand, the mass function
of model~3 (dot-dashed line) suppresses HI in larger halos, with the HI being
redistributed to lower mass halos.  

In the right-hand panel of Figure~\ref{fig_massfn}, we plot the mass
fraction of HI ($f_{\mathrm{HI}} = M_{\mathrm{HI}}/M_{\mathrm{halo}}$)
as a function of the mass of the host halo.  In all three cases, the HI
fraction is peaked around halos of mass in the range $6 \times 10^{10}
h^{-1}\msun < M_{\mathrm{halo}} < 2\times 10^{11} h^{-1}\msun$, and the
peak value is $14\%$ of the baryon mass fraction $f_{\mathrm{bar}} =
\Omega_b/\Omega_{\mathrm{m}} = 0.17$.  Lower and higher mass halos have
a suppressed HI fraction, which is due to the ratio of slopes $(m/n)$
and $(m/p)$ being smaller or equal to unity in Eqn.~(\ref{eq_MHI_M}).
The dependence of the HI mass on halo mass is again reflected by the
three models.  At lower halo masses, model~3 has a higher HI fraction
followed by model~1 and model~2, however, at the higher mass end the
situation is reversed.  This is also illustrated in
Figure~\ref{fig_slice}, where we show the distribution of dark matter
particles at $z = 1$ in a thin slice through our simulation enclosing
the largest halo.  The bottom panels zoom-in to a region of dimension
$50 \times 50 \times 4 \,(h^{-1}\rm{Mpc})^3$ centred on the largest
halo and showing the HI fraction $f_{\mathrm{HI}}$ for halos only, for
the three models (left to right).  As discussed earlier, the smaller
mass halos in model~3 have a higher neutral fraction in comparison to
the other two models.  In all models the largest halos have a smaller
neutral fraction, with model~2 dominating over the other two models.

Before proceeding, it is worthwhile to point out the advantages of using
a numerical N-body simulation over a halo-model based approach. In the
latter, properties like halo (and subhalo) abundances, halo profiles,
velocity dispersions and the halo-to-halo scatter are typically
calibrated from simulations such as ours (even so often based on much
smaller ones).  Once appropriate fitting functions are determined, they
can be used to predict the signal to a certain accuracy. However, the
approach cannot be used to construct a mock map with which the
efficiency of signal extracting can be studied.  The clustered
distribution of halos in a volume is crucial for properly describing
the signal.  Halos often occupy common pixels in a map and may add in
various combinations to the total signal of a given pixel.  The noise in
a pixel is a fixed random value irrespective of how many halos
contribute to the signal in that pixel.  Techniques for extracting the
signal from a map need to be explored when presenting results for its
detectability with instruments.  A halo model is generally not able to
account for these effects accurately and as a result tends to
overpredict the significance of detection.  This issue will become
clearer when we discuss how the signal is extracted from a mock map for
the noise levels that we consider, in Sections~\ref{sec_h1signal} and
\ref{sec_results}.

\subsection{A Common Field of View for DEEP2 and the GMRT} 

In this Section, we describe our fiducial choices for volume and halos
based on the specifications of the DEEP2 survey and the GMRT.  The GMRT
is a radio interferometer, consisting of thirty $45\,{\rm m}$ diameter
antennas spread over $25\,{\rm km}$. Half of the antennas are spread
over a central compact array of diameter $1\,{\rm km}$, and the
remaining half are spread on 3 arms of length $14\,{\rm km}$ in a
Y-shaped distribution. The longest baseline is $26\,{\rm km}$ and the
shortest $100\,{\rm m}$. The GMRT operates on 5 central frequencies
($151\,{\rm MHz}$, $235\,{\rm MHz}$, $325\,{\rm MHz}$, $610\,{\rm MHz}$,
$1420\,{\rm MHz}$). For this work our focus is on the $610\,{\rm MHz}$
frequency which corresponds to a redshift of $z=1.3$ for the $21\,{\rm
  cm}$ line.  The operational redshift at this frequency is $ 1.18 \leq
z \leq 1.44$.  The angular resolution (corresponding to the largest
effective baseline) is 5 arcsecs, this translates to a comoving scale of
$d = 114\, h^{-1}\mathrm{Mpc}$.  The system temperature is
$T_{\mathrm{sys}} = 102\,{\rm K}$ and the antenna sensitivity $K =
0.32$.  The GMRT has a full bandwidth of $32\,{\rm MHz}$ over 256
channels.

The DEEP2 survey is a redshift survey with spectra for $\simeq 40000$
galaxies in the redshift range $0.7 \leq z \leq 1.4$.  The survey covers
4 strips of dimensions $0.5^\circ \times 2^\circ$ of the sky which
corresponds to $20\times80\, {h}^{-1}\mathrm{Mpc}$ (comoving) at $z =
1$. The total comoving volume of DEEP2 is $6 \times 10^6
({h}^{-1}\mathrm{Mpc})^3$.  The spectral resolution of DEEP2 is $\simeq
68\,{\rm km\,s^{-1}}$, and targets were preselected to a limiting
magnitude of $\mathrm{R} = 24.1$.  The DEEP2 spectroscopically targets
$\simeq60\%$ of the objects that pass the apparent magnitude limit. We
hence take the completeness of DEEP2 to be $\simeq60\%$.

The overlapping redshift range between DEEP2 and GMRT is $1.18 \leq z
\leq 1.4$, which corresponds to $288\, {h}^{-1}\mathrm{Mpc}$ in depth,
or nearly a quarter of the DEEP2 volume.  Our HI model is at the fixed
redshift $z = 1$ of the simulation output, and not exactly matched to
the redshift range of the GMRT, which would require a light cone
simulation.  Our results should however be a good approximation to
redshift averaged quantities such as the mass function.  We choose our
analysis volume to be of dimension $50 \times 80 \times 400
(h^{-1}\mathrm{Mpc})^3$, which is a quarter of the DEEP2 volume.  In
order to match the required number density of galaxies in DEEP2 (and account
for its finite completeness), we choose a minimum threshold mass of $M >
10^{11.4} h^{-1}\msun$.  We have compared this threshold to that
computed using the known luminosities and estimated mass-to-light ratios
for the DEEP2 galaxies \citep{2007ApJ...654..153C} and find good
agreement.  Above this threshold mass, there are 16388 halos in the
subvolume which we identify as galaxies. We also adopt two larger
threshold masses for testing our detection strategy; these are $M >
10^{12.0} h^{-1}\msun$ and $M > 10^{12.5} h^{-1}\msun$. For these mass
cuts there are 3835 and 1031 halos, respectively.

\subsection{Comparison with Observations}
\label{sec_obs}

Recently, \cite{2010Natur.466..463C} reported the first detection of HI
in emission from $z \simeq 0.8$. They cross-correlated the optical
galaxy density field (from DEEP2) with the signal from the redshifted HI
line using the Green Bank Telescope (GBT) to obtain a $4\sigma$
detection.  At $z \simeq 0.8$, the GBT's angular resolution corresponds
to a FWHM of $9\, h^{-1} \, \rm Mpc$ (comoving), but the frequency
resolution $\simeq 2\, h^{-1} \, \rm Mpc$ is much finer. Owing to the
much poorer angular resolution, \cite{2010Natur.466..463C} computed the
cross correlation along the line of sight direction.

To make a detailed comparison with these observational results, we
convolve our simulation box with the angular and frequency resolution to
match the analysis of \cite{2010Natur.466..463C}. We also follow
\cite{2010Natur.466..463C} in assuming pixels of size
$(2\,h^{-1}\mathrm{Mpc})^3$.  Note that our density field is at $z
\simeq 1$ whereas the observations are at $z \simeq 0.8$.  This should
not pose a serious problem while comparing results, since the only
quantity which changes is the mass function of halos and this variation
can be absorbed within the models that we consider.  We first compute
the fluctuating component of both the HI and the galaxy density field on
a pixel of size $(2\, h^{-1}\mathrm{Mpc})^3$. We then convolve these two
fields with GBT's point spread function, modelled as a Gaussian with
FWHM of $9\, h^{-1}\mathrm{Mpc}$ in the transverse direction and a
top-hat of width $2\, h^{-1}\mathrm{Mpc}$ in the redshift direction.

In order to mimic the optical-HI observations, we only assign halos with
$M > 10^{11.4}\,h^{-1}\msun$ in the volume while constructing the galaxy
density field.  We do not use such a threshold when constructing the
HI density field.  The cross-correlation as a function of relative
displacement, $r_z$, along the line-of-sight direction, can be expressed as
\citep{2010Natur.466..463C}: \baq \xi_{\mathrm{HI,opt}}(r_z) &=& \langle
\Delta T_b(d+r_z) \delta_{\mathrm{opt}}(d)\rangle \nonumber\\ &=& 284
\mu \mathrm{K}
\langle\delta_{\mathrm{HI}}(d+r_z)\delta_{\mathrm{opt}}(d)\rangle
\left(\frac{\Omega_{\mathrm{HI}}}{10^{-3}}\right)
\left(\frac{h}{0.72}\right) \nonumber\\ &\times&
\left(\frac{\Omega_{\mathrm{m}} +
  (1+z)^{-3}\Omega_{\Lambda}}{0.37}\right)^{-0.5}
\left(\frac{1+z}{1.8}\right)^{0.5} .
\label{eq_crosscorr}
\eaq Here $T_b = 284\,\mu{\rm K}$ is the $21\,{\rm cm}$ mean sky
brightness temperature, $\delta_{\mathrm{opt}}$ is the optical density
field and $\delta_{\mathrm{HI}}$ is the neutral hydrogen density field.
They are related by $\delta_{\mathrm{HI}}=br\delta_{\mathrm{opt}}$,
where
$b=\langle\delta^2_{\mathrm{HI}}\rangle^{1/2}/\langle\delta^2_{\mathrm{opt}}\rangle^{1/2}$
is the bias and
$r=\langle\delta_{\mathrm{HI}}\delta_{\mathrm{opt}}\rangle/
\left(\langle\delta^2_{\mathrm{HI}}\rangle\langle\delta^2_{\mathrm{opt}}\rangle\right)^{1/2}$
is the stochasticity. By construction $|r| \leq 1$.  Inserting this into
Eqn.~(\ref{eq_crosscorr}), one can see that the amplitude of the cross
correlation function determines the degenerate combination
$br\Omega_{\mathrm{HI}}$.  \cite{2010Natur.466..463C} put a constraint
on this combination of parameters, obtaining $br\Omega_{\mathrm{HI}} =
(5.5\pm 1.5)\times10^{-4}$.  In our simulation we can break this
degeneracy.  Note that $r$ and $b$ are both dimensionless and do not
depend on $\Omega_{\mathrm{HI}}$.  We find for the three models
$b=(0.578,0.641,0.538)$ and $r=(0.923,0.945,0.916)$, respectively.
Using the smallest value of $rb$, we obtain a constraint on
$\Omega_{\mathrm{HI}}$ of the form $\Omega_{\mathrm{HI}} = (1.16 \pm
0.30)\times 10^{-3}$, which is consistent with the value of
$\Omega_{\mathrm{HI}} = 10^{-3}$ taken in our study.

\begin{figure}
\begin{tabular}{cc}
  \includegraphics[width=3.3truein]{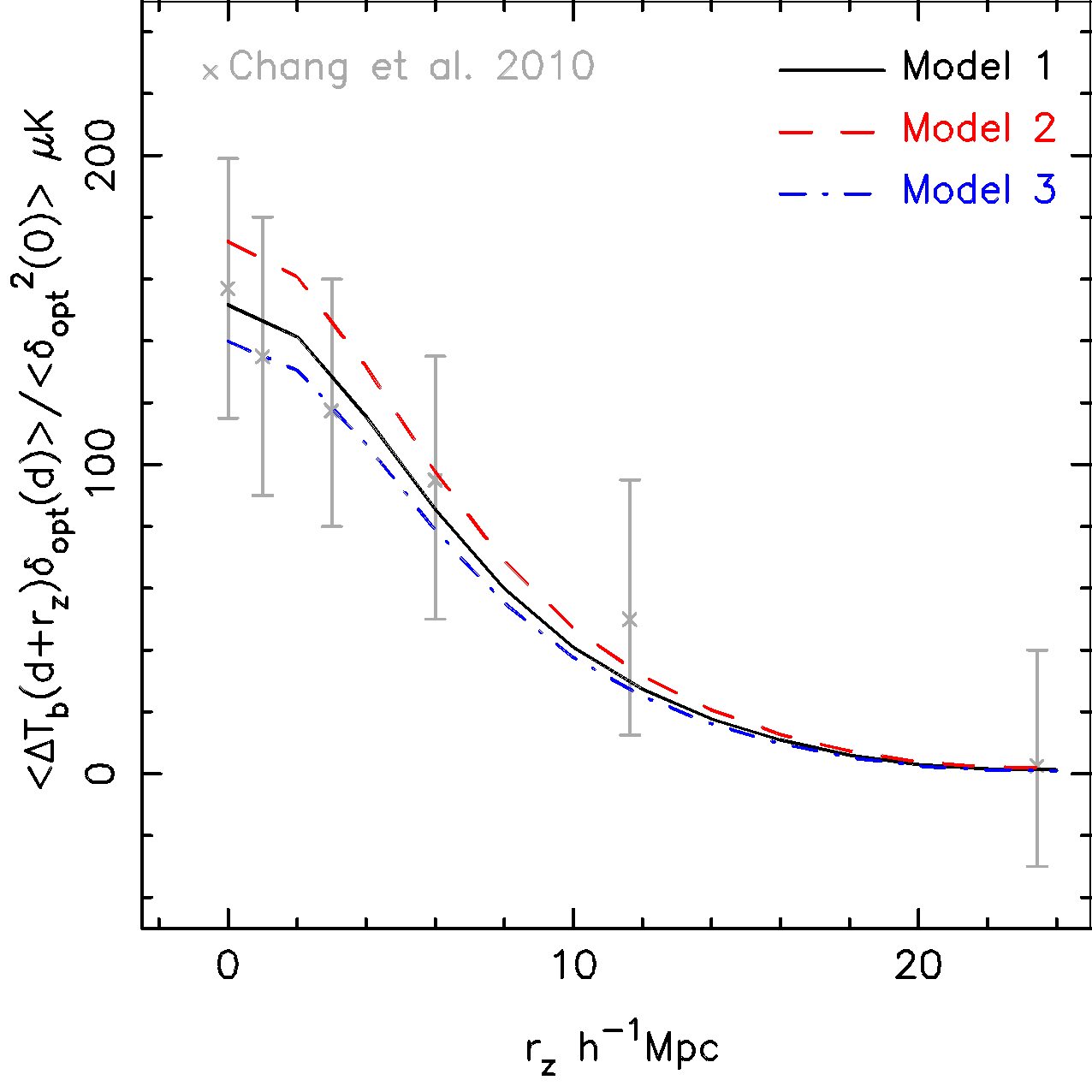} 
\end{tabular}
\caption{Normalised cross-correlation function of the DEEP2 optical
  galaxy density field and the HI intensity field along the
  line-of-sight direction \citep{2010Natur.466..463C} (data points), and
  the models 1 (solid line), 2 (dashed line) and 3 (dot-dashed line)
  that we consider.  Note that the cross-correlation function is
  normalised by the zero-lag auto-correlation function
  $\delta^2_{opt}(0)$ of the DEEP2 optical galaxy density field.}
\label{fig_crosscorr}
\end{figure}

\cite{2010Natur.466..463C} also computed the cross-correlation along the
line of sight direction and normalised it by
$\langle\delta^2_{\mathrm{opt}}(0)\rangle$.  We plot the normalised
cross-correlation function in Figure~\ref{fig_crosscorr} for the three
models (solid, dashed and dot-dashed) and compare with the observations
of \cite{2010Natur.466..463C}.  We find that all the three models are
consistent with the observations.  Note that model~2 is the most biased
of the three models, followed by models~1 and 3.  This is expected since
the largest halos have a considerably larger HI fraction in model~2, and
the largest halos cluster more strongly at smaller scales. A suppression
of HI in the largest halos will translate to a lower small-scale bias.
The large scale bias is further discussed in the following section.

\subsection{Finite Volume Effects on  $\Omega_{\mathrm{HI}}$}
\label{subsubsec_finvol}

\begin{figure}
\begin{tabular}{c}
\includegraphics[width=3.3truein]{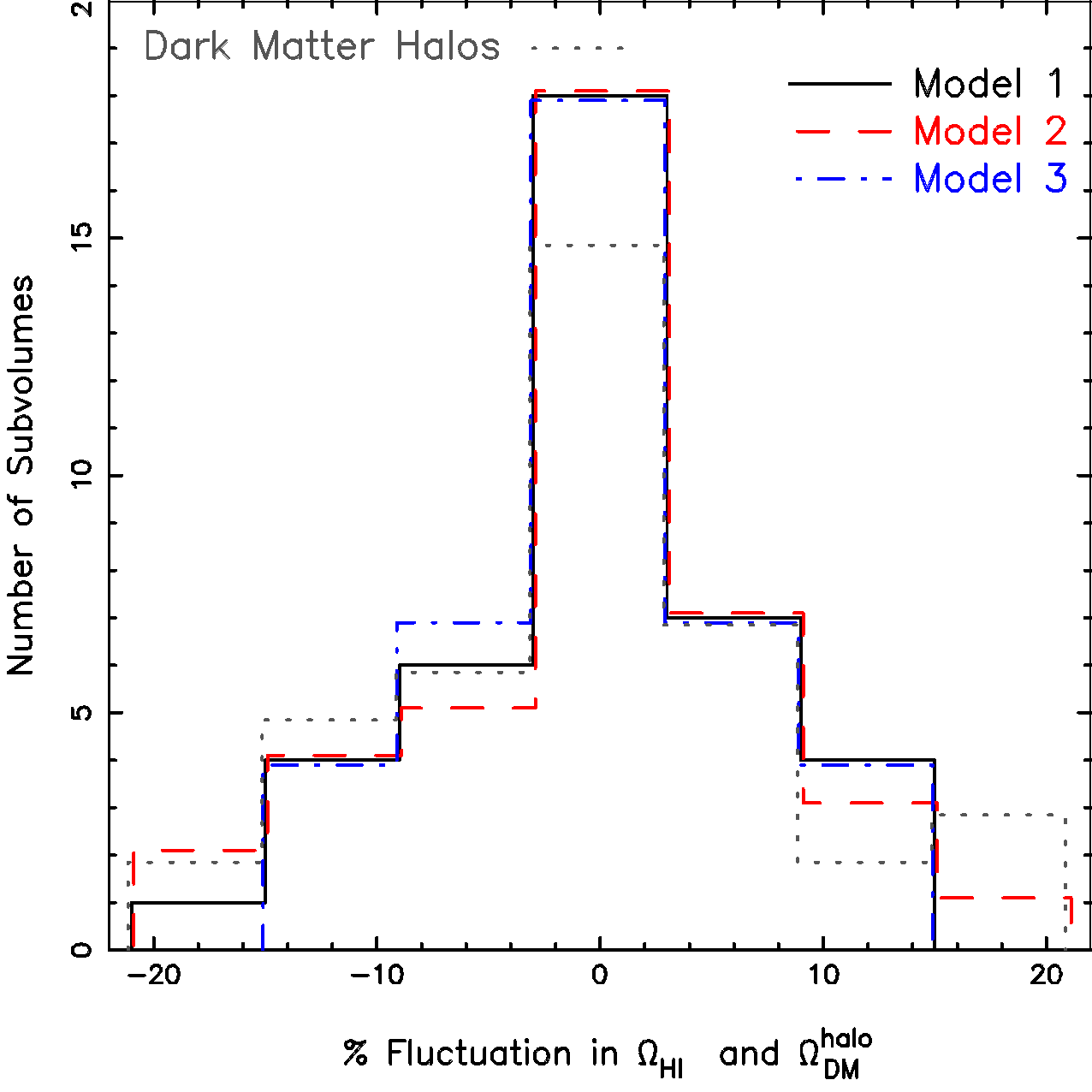} 
\end{tabular}
\caption{Fluctuations in $\Omega_{\mathrm{HI}}$ for the three models for
  a subvolume of dimension $50 \times 80 \times 400
  (h^{-1}\mathrm{Mpc})^3$ that we consider.  We have a total of 40 such
  subvolumes. The subvolumes considered for computing fluctuations in
  $\Omega_{\mathrm{HI}}$ were the same for all three models. The
  $1\sigma$ fluctuations for the three models are $6.8$\%, $7.6$\% and
  $6.2$\% respectively, for dark matter this value is $8.9$\%.}
\label{fig_cvar}
\end{figure}

In Figure~\ref{fig_cvar}, we look at finite volume effects in the
estimation of $\Omega_{\mathrm{HI}}$.  The size of the subvolume is
chosen so as to match the overlapping fields of DEEP2 and the GMRT.  In
our full simulation volume we have 40 such subvolumes.  We look at the
variations in HI mass in a subvolume with respect to the average HI mass
in the entire volume for the three models.  These variations are shown
for all three models in figure~\ref{fig_cvar} with the same line styles
as in Fig.~\ref{fig_massfn}.  We also plot the variation in
$\Omega_{\mathrm{DM}}^{\rm halo}$ (dotted line).  The rms fluctuation in
$\Omega_{\mathrm{HI}}$ for the three models is $\simeq6.8 \%$, $7.6\%$
and $6.2\%$ respectively, whereas the rms fluctuation in
$\Omega_{\mathrm{DM}}^{\rm halo}$ is $8.9\%$.  Given that the neutral
mass fraction is not uniform but rather peaked around halo masses in the
range of $6 \times 10^{10} h^{-1}\msun < M_{\mathrm{halo}} < 2\times
10^{11} h^{-1}\msun$ and suppressed for larger and smaller masses
(Fig.~\ref{fig_massfn}), the dark matter halos are more strongly biased
than HI.  This is consistent with \cite{2010MNRAS.407..567B} who showed
that the large scale HI bias increases with a higher neutral fraction in
larger halos.  Indeed we see that the fluctuation in
$\Omega_{\mathrm{HI}}$ in model 2, which has more HI in larger halos, is
closer to that of dark matter halos than the other two models, with
model 3 having the least fluctuations.  The final volume for our
analysis is picked based on the consideration that it should have the
smallest fluctuations in HI mass with respect to the mean for the
reference model~1.

\section{The 21 cm Emission Signal}
\label{sec_h1signal}

Since the $21\,{\rm cm}$ line has much larger wavelength than any
optical line, the resolution of a radio image is generally much poorer
compared to an image made in the optical.  Especially at high redshift,
a typical radio observation will be looking at most at a few coarse
pixels enclosing the target object rather than resolving it with a large
number of finer pixels.

In order to create a simulated radio data cube we create a mesh out of
the box with each pixel corresponding to an angular width of
$125\,h^{-1}{\rm kpc}$ (comoving), and frequency depth of $125\,{\rm
  kHz}$ (this is the width per channel of the GMRT for a single pointing
and matches the spectral resolution of DEEP2).  The angular resolution
is chosen to match that corresponding to the largest baseline of the
GMRT.  The HI mass in every pixel is computed by integrating the HI mass
profiles of halos in the pixels they cover, where a Gaussian profile with
a width given by the velocity dispersion of the halo is assumed in
redshift or frequency space.  Since observations are done in redshift
space, we have added the line-of-sight component of the peculiar velocity
of the halo to its real space line-of-sight $z$-coordinate to obtain its
redshift space coordinate.

The stacking of halos is done in the following manner.  Halos are first
sorted according to their mass.  We then identify the central pixel of
the halo corresponding to its redshift and its centre in the image
plane.  Given the location of halos as well as their angular and
frequency widths, we first select pixels along a line-of-sight (in
frequency) and passing through the central pixel.  Stacking is done on
the central or zero-reference frequency.  For every halo $i$ the
frequency range stacked is $\pm (4\times \Delta\nu_i)$ around the halo
centre.   Once stacked pixels are flagged so as to avoid double counting.
After this is done for every target halo, we repeat this procedure for
lines-of-sight not passing through the centre but neighbouring pixels in
cases where the halo is spread across more pixels.  Finally, the search
for pixels in frequency space is increased in order to stack the wings
of the signal.  

In case two target halos whose centres lie within the same line-of-sight
are overlapping within $\pm (4\times \Delta\nu)$ of each other, parts of
the smaller target halo may appear on the wings of the stacked spectra.
However, the order of stacking ensures that two halos along the same
line of sight are stacked in an optimal manner.  If we had chosen to
stack around the first halo with the entire frequency range
(corresponding to the box), then the second halo would appear on the
wing of the first halo.  We have checked that with a frequency width of
the pixel finer than $64\,{\rm kHz}$ we are able to recover the signal
reasonably well with this method, similar to what one would get from
just stacking analytically (as in a halo model) the flux, $S_{\nu}$, of
selected halos of mass $M_{\mathrm{HI}}$ located at a luminosity
distance $D_{\mathrm{L}}(z)$ with a line profile $\phi(\nu)$:  \beq
S_{\nu} = \frac{3}{4} \frac{A_{12} M_{\mathrm{HI}} h
  \nu}{m_{\mathrm{H}}} \frac{1+z}{4\pi
  D_{\mathrm{L}}(z)^2}\phi(\Delta\nu)
\label{eq_signal}
\eeq 
Here $\nu$ is the redshifted frequency $\nu = \nu_0/(1+z)$,
$A_{12}$ is the Einstein coefficient for spontaneous transition from the
upper to the lower level, $h$ is Planck's constant and $m_{\mathrm{H}}$
is the mass of the hydrogen atom. $\phi(\Delta\nu)$ is the line profile
which we take to be a Gaussian of width $\Delta\nu = \nu (\Delta v/c)$,
with $\Delta v$ being the velocity dispersion of the halo.

\subsection{Effect of Subhalos on the HI Signal}
\label{subsec_subhalos}

\begin{figure*}
\begin{tabular}{c}
\includegraphics[width=6.6truein]{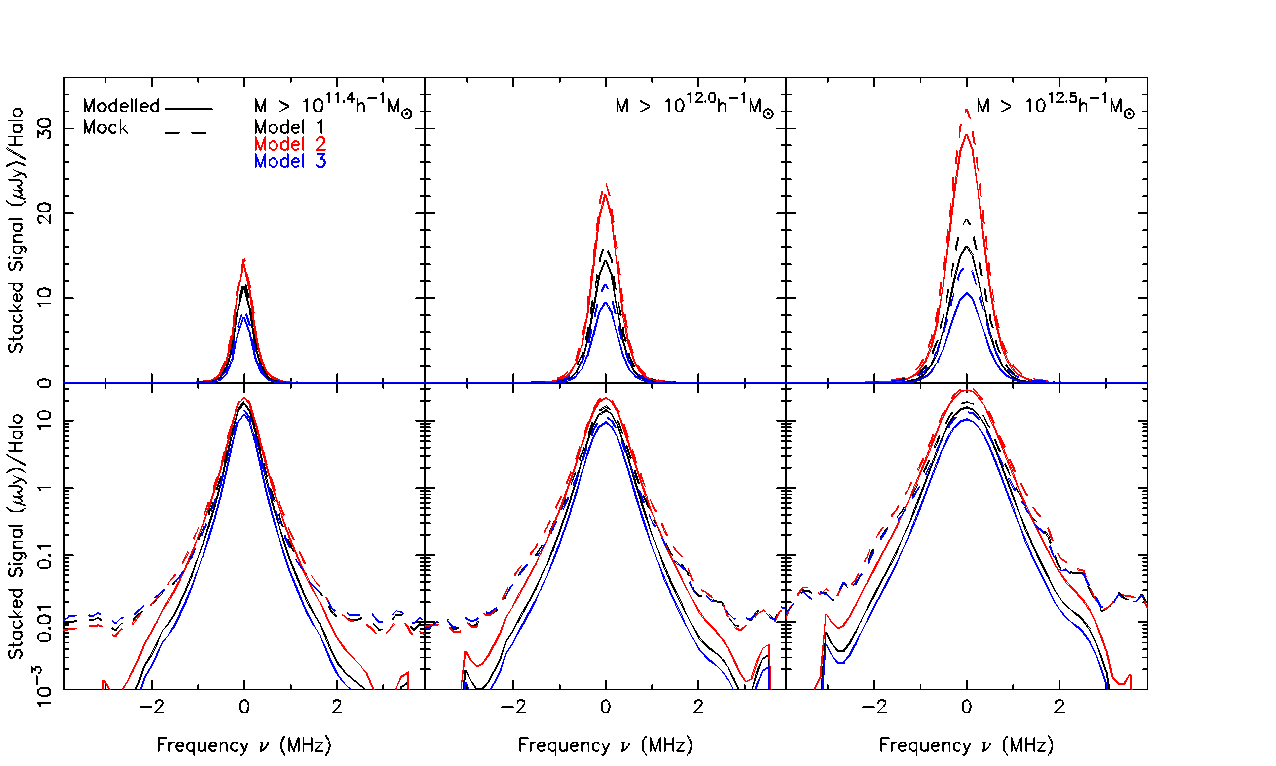}
\end{tabular}
\caption{The modelled HI emission signal per halo (solid) and the mock
  signal (dashed), recovered from the radio data cube by stacking, for
  the three mass cuts $M \geq 10^{11.4}h^{-1}\msun$ (left), $M \geq
  10^{12.0} h^{-1}\msun$ (centre), $M \geq 10^{12.5} h^{-1}\msun$
  (right).  The excess signal in the mock data is due to halos and
  subhalos below the mass thresholds, which were not identified for
  stacking.  The second row is the same as the first but replotted on a
  logarithmic $y$-axis to better illustrate the broader wings due to
  subhalos.}
\label{THEORYvsOBS}
\end{figure*}

The signal computed from a halo catalogue is different to that extracted
from a map.  The HI content in the pixels enclosing a target object is
typically greater than the HI mass of the object since lower mass halos
as well as interlopers (due to peculiar velocities) add to the pixel
with their own HI mass, thereby increasing the signal.  This effect is
larger in redshift space.  In Figure~\ref{THEORYvsOBS} we look at this
effect for all the models and for the three mass cuts of $M \geq
10^{11.4}h^{-1}\msun$ (left), $M \geq 10^{12.0}h^{-1}\msun$ (centre), $M
\geq 10^{12.5}h^{-1}\msun$ (right).  The average signal (in
$\mu\mathrm{Jy}$) per halo is plotted as a function of frequency, the
zero-frequency marks the central frequency where we have stacked spectra
of individual halos.  We compare the signal a mock observation would
measure (dashed line) when targeting objects with masses above a
threshold mass, with the theoretical or modelled expectation (solid
line).  The modelled signal was constructed by assigning only those
halos above the mass cuts in the data cube. Halos below the threshold
mass were not assigned to the data cube. In the mock observation, all
halos were assigned to the radio data cube and the spectra were stacked
for halos above the threshold mass.  The contribution of lower mass
halos can be seen in Figure~\ref{THEORYvsOBS} where the plots in the
second row are the same as those in the first row, but replotted on
log-y scale.  This is done so as to better illustrate the difference in
the wings of the stacked signal, with and without the subhalos.  The
average signal decreases with decreasing mass cut since lower mass halos
have on average a lower peak signal.  There is also scatter in the
relationship between halo mass (hence HI mass) and peak signal since
both the velocity dispersion and the HI mass determine the shape of the
signal.

In the mock spectra we find an enhanced peak and broader wings for all
the three mass cuts and models.  The enhanced signal being larger for
the larger mass cuts.  The enhancement is as much as $31\%$ for a mass
cut of $M \geq 10^{12.5}h^{-1}\msun$, decreasing to $9\%$ for a mass cut
of $M \geq 10^{11.4}h^{-1}\msun$ in model~3 where there is relatively
more HI in lower mass halos.  The numbers for model~2 are smallest where
the HI mass is dominated by larger mass halos, decreasing from $10\%$
for $M \geq 10^{12.5}h^{-1}\msun$ to $4\%$ for $M \geq
10^{11.4}h^{-1}\msun$.  For model~1, the intermediate model, the
contribution of subhalos is $22\%$ for $M \geq 10^{12.5}h^{-1}\msun$ and
decreases to $7\%$ for $M \geq 10^{11.4}h^{-1}\msun$.

\begin{figure*}
\begin{tabular}{c}
\includegraphics[width=6.6truein]{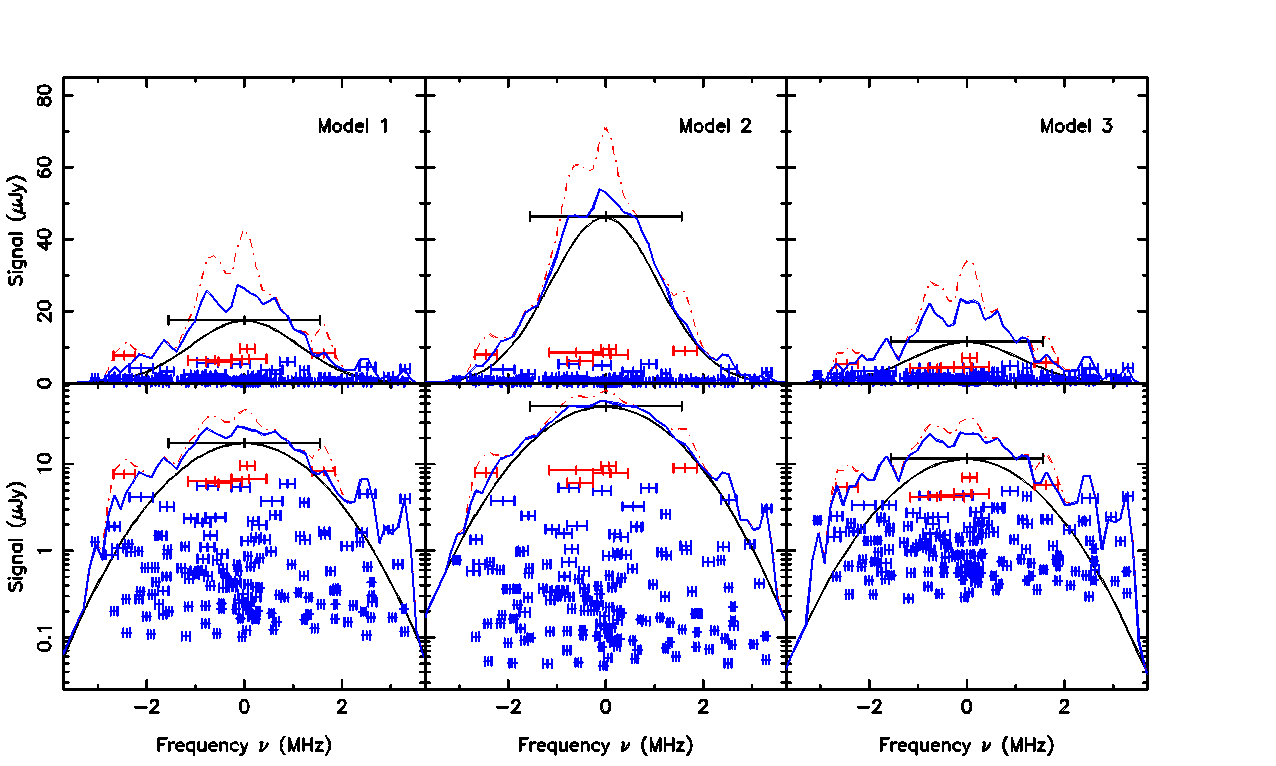} 
\end{tabular}
\caption{Contribution of subhalos and interlopers to the HI signal of a
  large halo for the three models (left to right).  An observation
  targeting the large halo would see an excess emission signal (blue
  solid line) compared to the expected (modelled) signal (black solid
  line) of the large halo.  Height and error-bars of each data point
  denote the peak and width of the theoretical signal of each subhalo or
  halo.  The excess signal is due to subhalos (blue data points) within
  the pixels of the targeted large halo; these are below the mass
  threshold of $M <10^{11.4} h^{-1}\msun$ and will not be identified in
  the optical survey.  The red data points are larger halos, also within
  the pixels covered by the targeted halo, which are above the mass
  threshold and would be identified in the optical survey.  If these
  were not separately picked for stacking then we would get an even
  larger signal (red dot-dashed line).  This particular halo has a mass
  of $M = 2\times 10^{14} h^{-1}\msun$.  The second row is the same as
  the first but replotted on a logarithmic $y$-axis to better illustrate
  the broader wings due to subhalos.}
\label{THEORYvsOBS_singleHALO}
\end{figure*}

The larger halos have more substructure as well as interlopers in
redshift space, both of which lead to an enhanced signal.  To illustrate
this point we pick one of the larger halos in the data cube and compute
the signal from the pixels that it covers. This is is shown in
Figure~\ref{THEORYvsOBS_singleHALO} for the three models (left to
right).  This particular halo has a mass of $M = 2 \times 10^{14}
h^{-1}\msun$.  The black solid line is the (theory) spectrum of the
large halo.  The blue data points are the theoretical spectra computed
from Eqn.~(\ref{eq_signal}) of all subhalos within the pixels covered by
the large halo, the height and error-bar being the peak signal and its
width.  These subhalos are below the mass threshold
$M<10^{11.4}h^{-1}\msun$ and will not be identified in an optical
survey.  An observation would see the emission (blue solid line) in
excess of the expected signal (black solid line) due to these subhalos.
This excess can be as much as $50\%$ of the total signal in model 3,
$13\%$ in model 2 and $35\%$ in model 1.  The red data points are larger
halos, also within the pixels covered by the targeted halo, which are
above the minimum mass threshold and would be identified in the optical
survey.  If these were not separately picked for stacking then we would
get an even larger signal (red dot-dashed line).  The enhanced signal
due to subhalos is within a $2\Delta\nu$ width of the large halo. If one
were to resolve all subhalos and stack them in this case then one would
get a peak signal in excess of $\simeq140\mu$Jy across models; instead
since these are unresolved within the pixel width and are spread across
the parent halo we get a peak signal in the range of $\simeq
22\hbox{--}55 \mu$Jy across models.  This effect will show up when we
constrain the HI mass function from the signal in a later section and
will boost the HI mass of the targeted halo.

One limitation of our model is that we assign an equal amount of HI to
both satellite and distinct field halos of the same mass. It is known
that gas is stripped from a halo when it merges into a larger halo.  Our
models hold if an equal fraction of cold gas and dark matter is stripped
from a halo during a merger.  \citet{2006ApJ...647..201C} argue that the
mass of a halo at the time of a merger, $M_{\mathrm{infall}}$ is a
better predictor of stellar mass (hence luminosity) than the mass of the
halo, $M_{\mathrm{obs}}$ when it is already a satellite.  By doing an
abundance matching of the luminosity function to the halo mass function
with their new definition of mass for satellites, their model reproduces
luminosity-dependant clustering of galaxies seen in
observations. However their new definition of mass of a satellite seems
to affect results more strongly at $z=0$ than at $z=1$ or higher.  If we
assume cold gas to trace stars and be more concentrated in the centre of
the halo, then based on the results of \citet{2006ApJ...647..201C} at
$z=1$, our model should not be sensitive to the environment of small
halos. However if cold gas is not concentrated in the centre of halos
and is largely stripped during a merger then our model may overpredict
the contribution of undetected satellites to the total signal of a large
halo.

\section{Results: Detecting HI in emission in the DEEP2 field with the GMRT}
\label{sec_results}
We now focus our attention to detecting HI in emission in the common
field of DEEP2 and the GMRT. We start with a discussion of noise in the
GMRT and then proceed to recover the stacked HI emission signal by
adding noise to the radio data cube.

\begin{figure*}
\begin{tabular}{c}
\includegraphics[width=6.6truein]{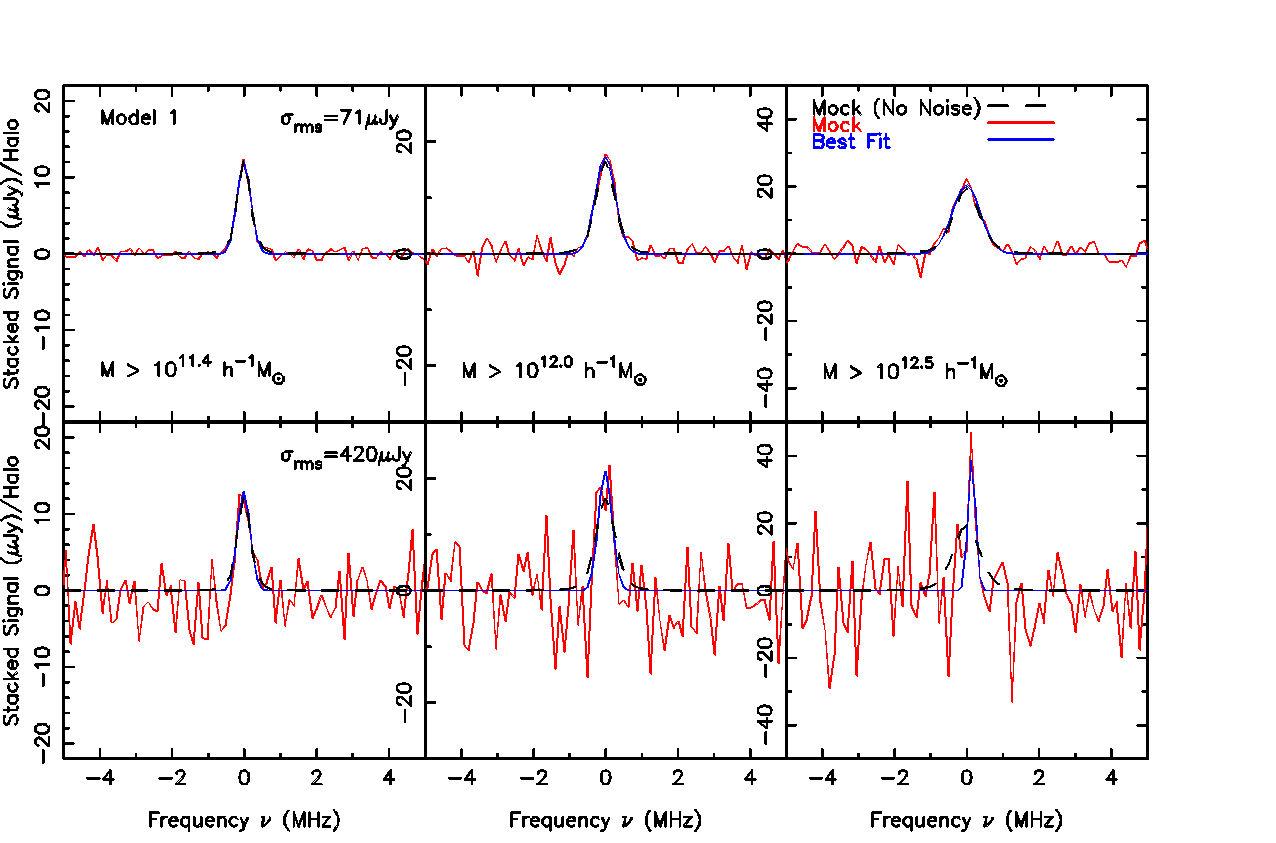}
\end{tabular}
\caption{Best fit spectra for model~1 by fitting a Gaussian to the mock
  spectra with rms noise of $71\,\mu {\rm Jy}$ (top) and $420\,\mu$Jy
  (bottom), for the three mass cuts $M \geq 10^{11.4}h^{-1}\msun$
  (left), $M \geq 10^{12.0}h^{-1}\msun$ (centre), $M \geq
  10^{12.5}h^{-1}\msun$ (right).  All three fits have
  $\chi^2_{\mathrm{red}} \simeq1$.}
\label{fig_signoisebf}
\end{figure*}

\subsection{Noise In Images}

The point source (or angular scales smaller than the synthesized beam of
the interferometer) sensitivity, $\sigma_{\rm rms}$, for an
interferometer is given by (assuming two polarizations)
\citep{2001isra.book.....T}:
\begin{equation}
\sigma_{\rm rms} = \frac{T_{\mathrm{sys}}}{K} \frac{1}{\sqrt{ \Delta \nu \Delta t}
  \sqrt{2\, N\left(N-1\right)}} 
\label{eqn_imagenoise}. 
\end{equation}
Where $K$ (in units of $\rm K/\mathrm{Jy}$) is the antenna sensitivity,
${T_{\mathrm{sys}}}$ is the system temperature and $N$ is the number of
antennas, $\Delta \nu$ is the channel bandwidth and $\Delta t$ is the
integration time.  The GMRT has a full bandwidth of $32\,{\rm MHz}$ with
256 channels, or $125\,{\rm kHz}$/channel for the maximum bandwidth in a
single pointing.  For this bandwidth and $N=30$, the noise per channel
is $\simeq71\mu$Jy for 24 hours of observation, where $K=0.32$ and
$T_{\mathrm{sys}}=102\,{\rm K}$ at the redshifts under consideration.

We need to make several other assumptions to bring the results of our
simulation closer to the possible observational outcome. For our
stacking approach, we need to co-add signal from sources occupying
different pixels in the three-dimensional data cube. However, the noise
in neighbouring pixels is not uncorrelated for a radio interferometer;
only the noise in different frequency channels is uncorrelated. To take 
this complication into account, we 
 assume here that the noise is uncorrelated for the spatially separated halos.
 However for all neighbouring
pixels enclosing a target object at a fixed frequency we choose the same
noise.  To take into account this uncertainty in estimating the noise
level and other complications owing to extraction of continuum point
sources, etc., we assume two different noise levels: $\sigma_{\rm rms} =
420 \, \rm \mu$Jy, which is an estimate of an upper limit, or
conservative, noise level on GMRT \cite[see e.g.][for a similar study at
  a neighbouring frequency]{2007MNRAS.376.1357L}, and $\sigma_{\rm rms}
= 71 \, \rm \mu$Jy, which corresponds to the theoretical (optimistic)
noise level computed for 24 hours of observation.

Both noise levels correspond to a pixel of size $125\,h^{-1}{\rm kpc}
\times 125\,h^{-1}{\rm kpc}$ (comoving), which is matched to the
approximate synthesized beam of GMRT at $\nu \simeq 700 \, \rm MHz$, and
depth $125\,{\rm kHz}$ for 24 hours of observation.  To every pixel we
add a Gaussian random noise with an rms of the two levels of noise that
we consider.

\subsection{Recovering the stacked HI emission spectra}

Having added noise to the radio data cube we attempt to recover the
stacked emission spectra by doing a $\chi^2$ analysis.  We model the
stacked spectra by a Gaussian with three parameters: \beq
S_{\mathrm{bf}} =
N_{\mathrm{bf}}\exp\left[-\left(\frac{\nu-\nu_{\mathrm{bf}}}{\Delta\nu_{\mathrm{bf}}}\right)^2\right]
\eeq where $N_{\mathrm{bf}}$, $\nu_{\mathrm{bf}}$,
$\Delta\nu_{\mathrm{bf}}$ are the best fit height, centre and width,
respectively.  We vary these three parameters over a large range and the
best fit values are obtained by minimizing $\chi^2$.  We illustrate this
analysis for the fiducial model~1 in Figure~\ref{fig_signoisebf} for the
three mass cuts (columns) discussed earlier and the two noise levels
(rows) we consider.  For a better illustration of the fits we have
changed the scales on the y-axis for the three mass cuts.  Note that
this is done for the mock spectra, where all subhalos have been added to
the radio data cube.  In all cases the reduced $\chi^2_{\mathrm{red}}
\simeq 1$ which is not a good indicator of the quality of the fit.  For
the optimistic noise (top row) we are able to correctly recover the
centre for all the three mass cuts.  Visually the best fit (solid blue)
seems to match the expected signal (black dashed) extremely well for
this case, but note that noise has not been included here. The red line
is is the mock spectra where noise has been added.  For the case of the
conservative noise, this is true for a mass cut of
$M>10^{11.4}h^{-1}\msun$ where we are able to recover the centre at the
zero reference frequency.  For a mass cut of $M>10^{12.0}\,h^{-1}\msun$
the best fit centre is not the zero centre but slightly shifted to the
left at $\nu_{\mathrm{bf}} = 25\,{\rm kHz}$.  For a mass cut of
$M>10^{12.5}h^{-1}\msun$, the best fit centre is incorrect and is
identified at $150\,{\rm kHz}$. We also find that the deviation from the
expected spectra is the largest for this case, where both the height and
the width of the spectra are considerably different than the expected
curve.  The quality of fits are similar for the other two models.

\begin{figure*}
\begin{tabular}{c}
\includegraphics[width=6.6truein]{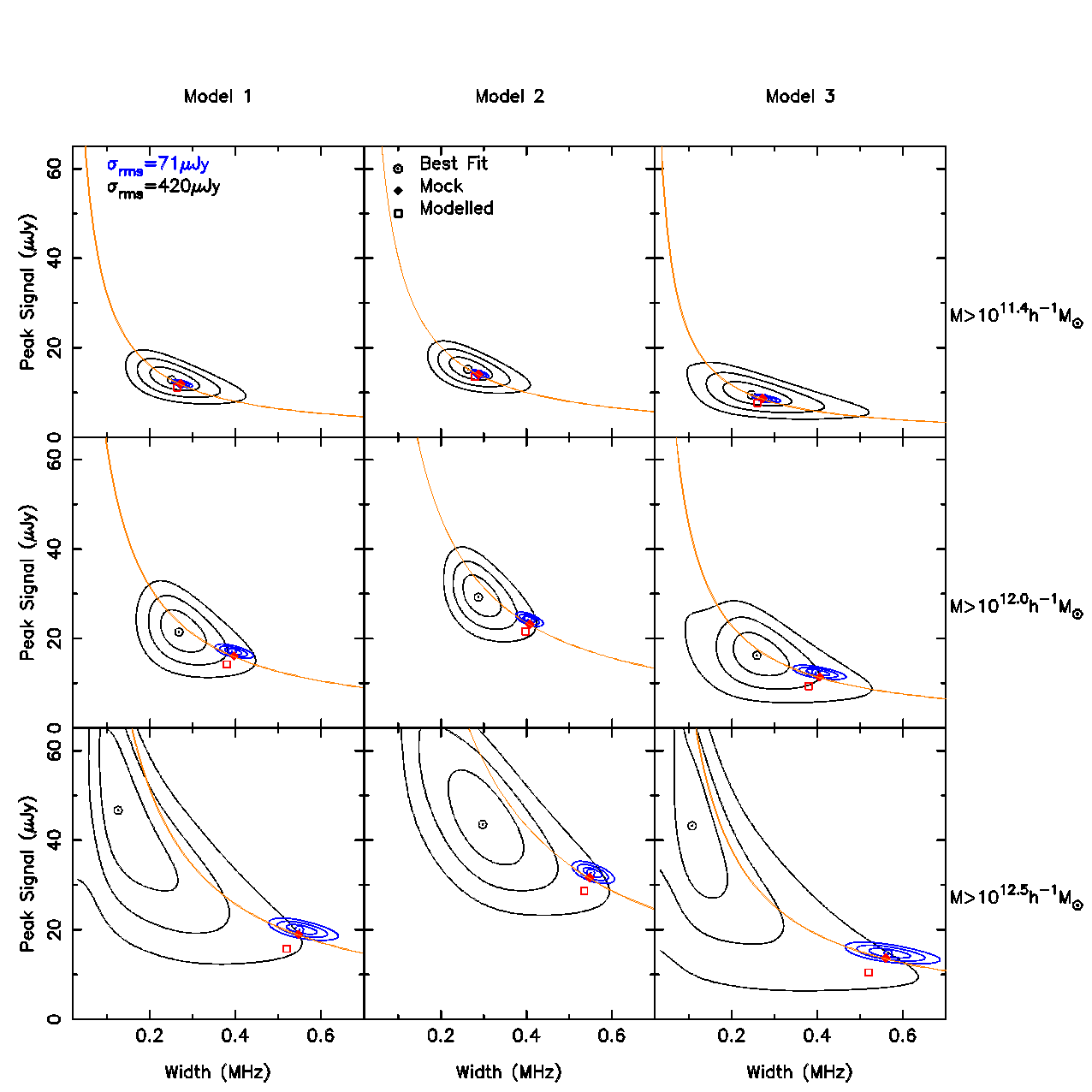}
\end{tabular}
\caption{Confidence contours for width and height of the fitted
  Gaussians for the three models (columns) and the three mass cuts $M
  \geq 10^{11.4} h^{-1}\msun$ , $M \geq 10^{12.0} h^{-1}\msun$ and $M
  \geq 10^{12.5} h^{-1}\msun$ (rows). The $1\sigma$, $2\sigma$ and
  $3\sigma$ contours are shown for both the conservative noise of $420\,
  \mu$Jy (black) and for the optimistic noise of $71\,\mu$Jy (blue).
  The dotted circles are the best-fit values from the mock spectra.  The
  open square is the expected point without subhalos, and the diamond is
  the expected point with subhalos, i.e.~mock spectra without noise.
  The solid ochre line shows the combinations of height and width which
  give the same mass as the expected average HI mass of the halo from
  the mock spectra.}
\label{fig_chisq}
\end{figure*}

We now move on to quantify the quality of the recovered spectra by doing
a likelihood analysis, where we marginalize over the centre,
$\nu_{\mathrm{bf}}$ and plot the $1\sigma$, $2\sigma$ and $3\sigma$
contours for the remaining two parameters, the width
$\Delta\nu_{\mathrm{bf}}$ and the height $N_{\mathrm{bf}}$ of the
stacked spectra. This is shown in Figure~\ref{fig_chisq}.  The columns
represent the models and the rows are for the three mass cuts.  The
dotted circle is the best-fit value of the width and height.  The black
contours are for the conservative noise of $420\,\mu$Jy and the blue
contours are for the optimistic noise of $71\,\mu$Jy.  The filled diamond
is the expected value of the mock spectra without noise, whereas the
open square is the same without subhalos. In all cases, the width and
height for the spectra without subhalos is smaller than for the ones with
subhalos, as was discussed in section~\ref{subsec_subhalos}. This is
shown again for reference.  For both noise levels, we see that the
contours are oriented in a manner showing an anti-correlation between
height and width.  This is expected since the product of the two
determines the mass of the object.  This degeneracy which determines the
mass of the object is shown in the solid ochre line passing through the
expected value of the mock spectra. An incorrect combination of the two
would give the same average HI mass per halo.

We find that for the optimistic noise the quality of the fit is
extremely good and the best fit values are within $1\sigma$ of the
expected values for the smallest mass threshold of $M >
10^{11.4} h^{-1}\msun$ and is within $3\sigma$ for the other two
mass cuts.  However, satellites below the threshold mass and contributing
to the HI mass of the target halo can be more strongly discriminated
with the larger mass cut.  The difference being the largest for model 3
and the least for model 2 as discussed in section~\ref{subsec_subhalos}.
We indeed find that the stacked spectra with and without features of
satellite galaxies can be discriminated by more than $7\sigma$ for the
two larger mass cuts and by $4\sigma$ for the smaller mass cut, in our
model.  The contribution of satellites in the lowest mass cut
corresponds to objects missing in the optical survey.

The GMRT will therefore be sensitive to subhalos that are undetected in
the optical survey (although optical stacking would be able to detect
them). Inferring what their fractional contribution is could be carried
out in various model dependent ways. One can look at specific model
predictions (as has been done in this paper) to compare to the stacked
signal.  One could instead determine how satellite galaxies populate the
central halo, i.e.  measure the halo occupation distribution, from a
different approach.  The latter approach should be feasible in a
statistical 21cm survey where one observes the 21cm power spectrum (or
the correlation function) out to small non-linear scales and infers the
HOD from it \citep{2010MNRAS.404..876W}.  This was done with the
correlation function of the HIPASS galaxies at $z \simeq0$
\citep{2009arXiv0908.2854W}.  Recently \cite{2010MNRAS.407..567B} showed
that a standalone statistical detection of 21cm clustering would not be
feasible with the GMRT or the MWA and will have to wait for future
instruments.

For the conservative noise of $420\,\mu$Jy, the best fit and the
expected value lie within 1$\sigma$ for a mass cut of $M >
10^{11.4}h^{-1}\msun$.  This degrades to 2$\sigma$ and 3$\sigma$ for the
larger mass cuts.  In this case, since the noise is larger, the
error-contours are also broader compared to a noise level of
$71\mu$Jy. We also find that the best fit value systematically veers off
the line of constant mass in the direction of lower mass as the
threshold mass is increased.  This trend is also seen for the lower
noise but in the direction of higher mass, but is less prominent.  The
realisation of noise decides in which direction the best fit value moves
when noise becomes important, since it may underpredict or overpredict
the HI mass.  We cannot distinguish the effect of satellites on the
spectra with the conservative noise, unlike the optimistic case.
However, as mentioned before, since the best fit value for the two lower
mass cuts still lie near the line of constant mass, we would not get the
cumulative HI mass wrong, even though the shape of the spectra differs
from the true shape for $M > 10^{12.5}h^{-1}\msun$ as seen in the bottom
right panel of Figure~\ref{fig_signoisebf}.

The rms fluctuations in the shape of the average spectra when
considering all the subvolumes are a few percent and below the
fluctuations in total mass.  This happens because we fit the average
spectra of halos above the mass threshold and not the total spectra.
The number of halos above a certain mass cut fluctuates more strongly as
an increasing function of this mass cut. Therefore, the effect of cosmic
variance is larger in the mass function as compared to the average
spectra. We will revisit the issue of cosmic variance on the estimates
of the HI mass function in the next section.  We do not plot the errors
due to cosmic variance in Figure~\ref{fig_chisq} since they are smaller
than the size of the symbols.

\subsection{Subsamples and Constraints on The HI Mass Function}

We now discuss the extent to which the HI mass function can be
constrained with the GMRT and DEEP2.  To obtain the HI mass per halo, or
the cumulative HI mass, we need to invert Eqn.~(\ref{eq_signal}).  We
assume a mean redshift $\bar{z}$ and a mean luminosity distance
$\bar{D}_{\mathrm{L}}$ of our survey, which we take to be at the centre
of our subvolume along the redshift direction.  The total HI mass is
then proportional to the height and the width of the fitted Gaussian and
the number of halos above the mass threshold.  The error in the HI mass
is hence dependant on both: the errors on the height and the width.  To
obtain the error on either height or width, we further marginalise our
likelihood function over the other parameter and compute the 1$\sigma$
errors on them.

\begin{figure*}
\begin{tabular}{c}
\includegraphics[width=6.2truein]{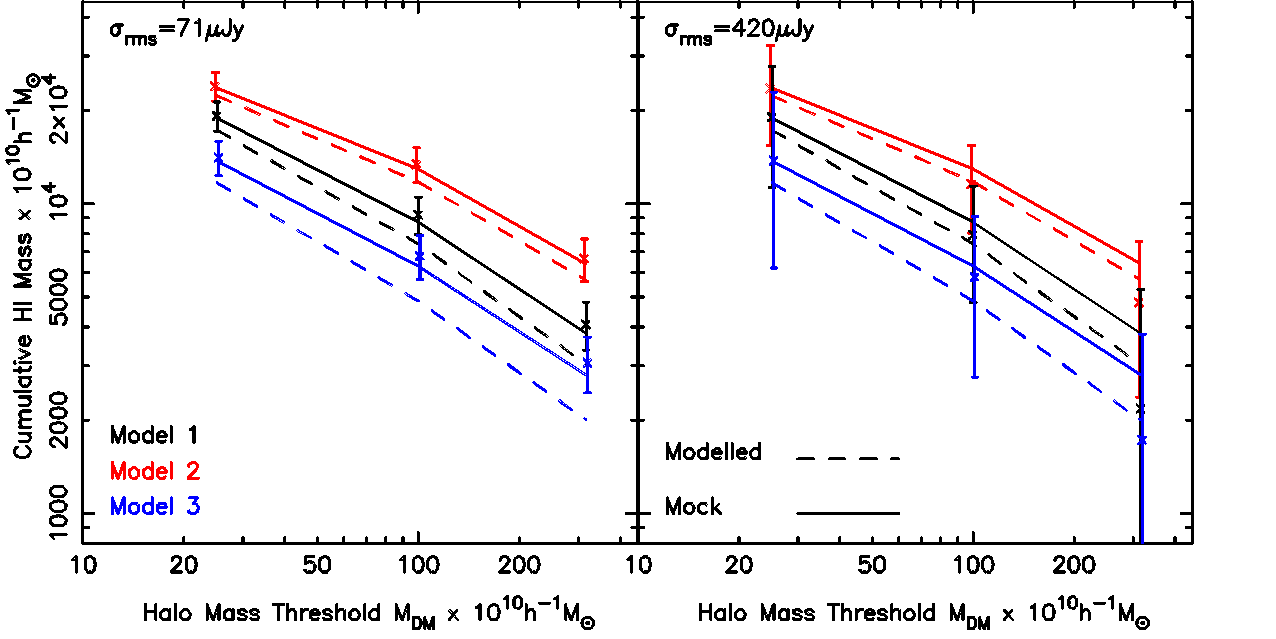}
\end{tabular}
\caption{The recovered cumulative HI mass function for the three models
  with the optimistic noise of $\sigma_{\rm rms} = 71\,\mu$Jy (left) and
  the conservative noise of $\sigma_{\rm rms} = 420\,\mu$Jy (right). The
  expected mass function with subhalos (solid line) and without subhalos
  (dashed line) are also drawn for comparison.  Data points were
  computed from the mock spectra.}
\label{fig_cummassfn}
\end{figure*}

We present the constraints on the cumulative HI mass function in
Figure~\ref{fig_cummassfn} for both the optimistic noise of $71\mu$Jy
(left) and the conservative noise of $420\mu$Jy (right).  The total HI
mass for halos above the cutoff mass of the halo has been plotted as a
function of cutoff mass of the halo.  The uncertainty of the best fit
parameters due to noise as well as fluctuations due to cosmic variance
have both been included in the error bars, and were added in quadrature.
The contribution of each is shown in Table~\ref{table_err}.  The solid
line is the expected cumulative HI mass function and the dashed line is
the same without satellites.

The effect of cosmic variance should be more pronounced for rarer or
more massive objects.  This is indeed the case, as is seen in
Table~\ref{table_err}, where we find that the fluctuations due to cosmic
variance increase with increasing threshold mass for all the three
models, the effect being largest for model~2 followed by model~1 and
model~3, consistent with the discussion in
section~\ref{subsubsec_finvol}.

In the optimistic case, the mass function can be well constrained over
the entire range of masses that we consider. Note that in this case the
best-fit points lie systematically above the mock HI mass function,
which can be attributed to the noise, as we discussed in the previous
section.  We had seen in Figure~\ref{fig_chisq} that the contribution of
satellites even for the lowest mass cut could be distinguished at the
$3\sigma$ level when we look at the stacked spectra.  This is not the
case for the mass function, where the cosmic variance is often more
dominant than the errors due to noise. We find that the modelled HI mass
function is within $1\sigma$ of the best fit mass function with
satellites for model~2 over the entire mass range.  This is not so for
model~3, where the modelled mass function is well beyond the $1\sigma$
from the mock mass function over the entire mass range that we consider
since the effect of satellites is more pronounced.  In model~1, the
modelled mass function is within the $1\sigma$ errors of the mock mass
function for $M > 10^{11.4}h^{-1}\msun$ and beyond it for the larger
mass cuts.  From Table~\ref{table_err}, we compute the detection
significance (including cosmic variance) for the optimistic noise.  We
find it is the highest for model 2, being $9.4\sigma$ for $M > 10^{11.4}
h^{-1}\msun$ and $11.6\sigma$ for $ M > 10^{12.5} h^{-1}\msun$.  For
models 1 and 3 the numbers are $(9.3\sigma, 5.6\sigma)$ and $(7.9\sigma,
4.9\sigma)$ respectively.

We now move on to the case of the conservative noise in the right panel
of Figure~\ref{fig_cummassfn}.  The best-fit points lie systematically
below the mock mass function, in this case, due to the different
realisation of noise than in the optimistic case.  Here the
uncertainties due to noise are much larger than those due to cosmic
variance.  The modelled HI mass function is well within $1\sigma$ of the
mock mass function, hence the effect of satellites cannot be seen.  We
find a $1.7\hbox{--}3\sigma$ detection for mass cuts in the range
$10^{11.4} h^{-1}\msun < M <10^{12.0}h^{-1}\msun$. A detection is not a
possible for the larger mass cut of $M > 10^{12.5}h^{-1}\msun$ for
models 1 and 3, whereas a weak detection is possible for model 2 for
this mass cut.

\begin{table}
      \begin{center}
        \begin{tabular}{c|l|l|l}
          \hline
          Errors & Model 1 & Model 2 & Model 3\\
          \hline
          $\sigma_{\rm cosm}(M>M_{11.4})$ & 8.63\% & 9.30\% & 8.59\% \\
          $\sigma_{\rm cosm}(M>M_{12.0})$ & 10.85\% & 11.43\% & 10.87\% \\
          $\sigma_{\rm cosm}(M>M_{12.5})$ & 13.56\% & 14.10\% & 13.58\% \\
         \hline
         \hline
          $\sigma_{\rm rms}=71\mu$Jy & & & \\
         \hline
          $\sigma_{_{M_{\mathrm{HI}}}}(M>M_{11.4})$ & 6.47\% & 5.23\% & 9.37\% \\
          $\sigma_{_{M_{\mathrm{HI}}}}(M>M_{12.0})$ & 8.55\% & 5.91\% & 12.11\% \\
          $\sigma_{_{M_{\mathrm{HI}}}}(M>M_{12.5})$ & 11.41\% & 6.87\% & 15.34\% \\
         \hline
         \hline
          $\sigma_{\rm rms}=420\mu$Jy & & & \\
         \hline
          $\sigma_{_{M_{\mathrm{HI}}}}(M>M_{11.4})$ & 42.23\% & 34.81\% & 60.04\% \\
          $\sigma_{_{M_{\mathrm{HI}}}}(M>M_{12.0})$ & 40.18\% & 29.82\% & 53.62\% \\
          $\sigma_{_{M_{\mathrm{HI}}}}(M>M_{12.5})$ & 108.97\% & 52.05\% & 135.63\% \\
         \hline
         \hline
        \end{tabular}
      \end{center}
\caption{Breakup of errors on the HI mass function due to cosmic
  variance and noise for models (columns) and the three mass cuts that
  we consider. The first three rows are the \% fluctuations due to
  cosmic variance for the three mass cuts. Due to lack of space we
  change the notation of masses, e.g.~$M_{11.4} \equiv 10^{11.4}
  h^{-1}\msun$. The next three (filled) rows are errors in mass
  estimates due to the optimistic noise of $71\,\mu$Jy and the final
  three (filled) rows are for errors in mass estimates for the
  conservative noise of $420\,\mu$Jy.}
\label{table_err}
\end{table}

\section{Discussion and Conclusions}
\label{sec_conclusions}

In this paper, we have studied the prospects for detecting HI in
emission at $z \simeq 1$. This is a crucial epoch in the study of galaxy
formation, since the cosmic star formation rate starts to decline around
this time and the missing link in observations is an accurate census of
cold gas, which fuels star formation, at these redshifts and beyond
\cite[for more discussion of the importance of this issue see
  e.g.][]{2009astro2010S.241P}.  We make a case that an existing
instrument like the GMRT can put strong constraints on the amount of
cold gas contained in galaxies, when it is combined with a survey like
DEEP2.  In this work, we have only focused on the overlapping volume of
DEEP2 and GMRT, which represents a quarter of the total DEEP2 volume.
Our study is representative of what might be achievable by combining the
already existing optical data and the presently operational radio
interferometers.

The HI signal is too weak in emission for the detection of individual
objects at $z\simeq1$. However, this can be circumvented by a stacking
strategy, similar to \cite{2007MNRAS.376.1357L,2009MNRAS.399.1447L},
which we here apply to look at the prospects of detection.  Our
conclusions are:

\begin{itemize}
    \item We find that a detection of HI in emission at redshifts of $z
      \simeq 1$ is possible even with existing instruments like the GMRT
      when combined with the DEEP2 survey. Such an observation will be
      able to constrain the HI mass function in the halo mass range
      $10^{11.4} h^{-1}\msun \leq M \leq 10^{12.5} h^{-1}\msun$.  The
      detection significance is in the range of $5\hbox{--}12\sigma$ for
      an optimistic noise level of $71\mu$Jy with 24 hours of
      integration.

  \item The models that we consider are consistent with recent
    observations of \cite{2010Natur.466..463C}, who computed the
    cross-correlation of the density field of DEEP2 galaxies and the
    21cm intensity field with the GBT. However these observations allow
    for all the three models that we consider. On the other hand, we
    find that using the stacking technique it will be possible to
    discriminate between the different scenarios with an instrument
    like the GMRT, at least for the optimistic level of noise.

  \item Combining our estimates of HI bias with the observations of
    \cite{2010Natur.466..463C}, we find that the most conservative
    constraint on the cosmic HI fraction at $z \simeq 0.8$ to be
    $\Omega_{\mathrm{HI}} = (1.16 \pm 0.30)\times 10^{-3}$.

  \item We find that undetected satellites in the optical produce a
    non-negligible contribution to the stacked HI spectra.  Their
    signature is better seen in the stacked spectra rather than in the
    mass function, since we integrate over one parameter, i.e.~the width
    of the spectra, to obtain the mass function.  For a noise of $71 \mu
    \mathrm{Jy}$, features of satellites can be seen at the $4\sigma$
    level in the stacked spectra for a mass threshold of $M \geq
    10^{11.4}h^{-1}\msun$.  This detection significance for satellites
    increases by more than $7\sigma$ for $M \geq 10^{12.0} h^{-1}\msun$
    (see e.g. Fig~\ref{fig_chisq} and Fig.~\ref{fig_cummassfn}).  In
    comparison, the mass function discriminates satellites at the $\sim
    1\sigma$ level.
    
  \item 
    We have also considered a much higher level of noise, i.e.~$420 \mu
    \mathrm{Jy}$, which should represent an upper bound on noise in the
    GMRT.  With this amount of noise, a detection of the mass function
    is possible at the $1.7\hbox{--}3\sigma$ level. We expect that
    the real detection significance is bracketed by our the optimistic
    and conservative noise levels.

  \item For the higher noise, the effect of satellites on the stacked
    spectra can be seen only at the $1\hbox{--}3\sigma$ level across the
    ranges of mass that we consider.  The best-fit parameters of the
    spectra however are incorrect for the larger mass cuts when
    compared to the theoretical numbers.

  \item Cosmic variance affects the mass function more strongly than the
    average stacked spectra. For this reason, if HI is populated in halos
    according to model 2 one cannot quantify the effect of subhalos on
    the mass function due to the effect of cosmic variance. For models 1
    and 3, cosmic variance does not swamp the errors due to noise.
    
\end{itemize}

One can use the stacking strategy to independently probe
$\Omega_{\mathrm{HI}}$ \citep{2009MNRAS.399.1447L}. This is not the case
in the cross-correlation approach which constrains
$br\Omega_{\mathrm{HI}}$.  As in \cite{2007MNRAS.376.1357L}, it would be
useful to also target a subset of galaxies in DEEP2 whose SFR has been
measured. This would provide the link between the SFR and the amount of
cold gas in galaxies and provide insight into models of galaxy
formation.  Since spectroscopic surveys are accurate but expensive it
would be worthwhile to first try this stacking strategy on future
surveys like the LSST, which are designed to give photometric redshifts
of $\simeq 10^{10}$ galaxies.  Photometric redshifts are more prone to
errors, but it has to be seen if the larger sample of a photo-$z$ survey
like LSST could beat down the noise by its sheer number of objects.

In this study, we have modelled the HI in all the halos, centrals and
satellites, and we have seen how the satellite population affects the HI
mass function as well as the stacked HI profile. The possibility to see
the effect of satellites missing in an optical survey in the
corresponding $21\,{\rm cm}$ survey is an exciting prospect.  On the one
hand, we find that stacking can distinguish between models, but the
effect of satellites on the stacked profile is model dependant, and to
see their effect one may need to combine it with the cross-correlation
approach.  In the cross-correlation method the optical density field
does not contain all the satellites, whereas the HI intensity field
does.  If we use the same mass threshold when constructing the HI
intensity field, one naively expects a stronger cross-correlation
between the two fields.  A preliminary investigation shows that this is
indeed the case.  We also expect that the HI bias and stochasticity will
be sensitive to subhalos.  A combination of both approaches would shed
light on both the model and the contribution of satellites.

The other approach is to observe the auto-correlation function or the
power spectrum of HI, and to constrain the HOD of HI galaxies
\citep{2010MNRAS.404..876W} from it.  Such an inferred model of HOD when
combined with a direct detection as is done here could reveal the
contribution of the satellite population on the total signal.  We will
look into these aspects of the analysis in a forthcoming paper.

Currently operational radio instruments -- both single dish and
interferometers -- have the capability to detect HI in emission at $z
\simeq 1$, as already demonstrated by \cite{2010Natur.466..463C}. We
explored the potential of these complementary strategies. In particular,
we studied in detail the efficiency of stacking, possible only with
interferometers.  In the near future, we expect larger optical galaxy
samples at $z\simeq 1$ and radio observations with wider field-of-views
and spectral coverage using upcoming radio instruments
\cite[e.g.][]{2007PASA...24..174J}.  This observational progress will
enable a better determination of the HI signal using either of the
strategies, thereby substantially improving our estimate of the HI
content of galaxies at $z \simeq 1$.
 
\section*{Acknowledgments}
We would like to thank Kevin Bandura and Jeff Peterson for useful
discussions on the analysis of their recent paper.  
We would like to thank Jeffrey A. Newman for useful discussions.
This work was supported by NSF award OCI-0749212. This research was supported by an
allocation of advanced computing resources provided by the National
Science Foundation.  The computations were performed on Kraken, Athena,
or Nautilus at the National Institute for Computational Sciences
(http://www.nics.tennessee.edu).

\label{lastpage}

\end{document}